\documentclass{aa}  
\def\apjl{ApJ}%
\def\apjs{ApJS}%
\def\ao{Appl.~Opt.}%
\def\aap{A\&A}%
\usepackage{graphicx}
\usepackage{subfigure}
\usepackage{natbib}
\bibpunct{(}{)}{;}{a}{}{,} 
\usepackage{txfonts}

\def\crh{count rate histogram}
\def\fin{Fin/Fout}

\def\redc{$\chi^{2}_{\mu}$}
\def\cratio{$R_{\chi}$}

\begin{document}
\title{Identifying XMM-Newton observations affected by solar wind
  charge exchange - Part I\thanks{Based on observations with XMM-{\it
      Newton}, an ESA Science Mission with instruments and
    contributions directly funded by ESA Member States and the USA
    (NASA).}}

   \subtitle{}

   \author{J.A. Carter
          \inst{1}
          \and
          S. Sembay\inst{1}
          }

   \offprints{J.A. Carter}

   \institute{Department of Physics and Astronomy, University of Leicester, 
 Leicester, LE1 1RH, UK\\
              \email{jac48@star.le.ac.uk}\\
	      \email{sfs5@star.le.ac.uk}\\
                }

   \date{Received ; accepted }

 
  \abstract
   {}
   {We describe a method for identifying XMM-Newton observations that
     have been affected by Solar Wind Charge Exchange (SWCX) emission
     and present preliminary results of previously unidentified cases
     of such emission within the XMM-Newton Science Archive.}
   {The method is based on detecting temporal variability in the
     diffuse X-ray background. We judge the variability of a low
     energy band, taken to represent the key indicators of charge
     exchange emission. We compare this to the variability of a
     continuum band, which is expected to be non-varying, even in the
     case when SWCX enhancement has occurred.}
     {We discuss previously published results with SWCX contamination
       that have been tested with the above method. We present a
       selection of observations that we consider to show previously
       unpublished SWCX enhancements, and further investigate these
       observations for correlation with data from the solar wind
       observatory, ACE. We also consider the geometry and viewing
       angle of XMM-Newton at the time of the observation to examine
       the origin of the charge exchange emission, whether it be from
       interactions with geocoronal neutrals in Earth's magnetosheath
       or from the heliosphere and heliopause.}
   {}


   \titlerunning{Identifying XMM-Newton obsns affected by SWCX - Part I}
   \authorrunning{J.A. Carter \& S. Sembay}
   \maketitle
%

\section{Introduction}\label{secintro}
This paper is the first of two articles which describe a new method
developed to search for instances of Solar Wind Charge Exchange (SWCX)
contamination of XMM-Newton observations. In this paper we outline the
method and provide an initial sample of results. In paper II we will
present the results of applying this method to all suitable
observations within the public XMM-Newton Science Archive (XSA).

In this introduction we briefly discuss the SWCX process including
examples where this effect occurs in the solar system, and then
describe the XMM-Newton observatory and how its orbit affects the
likelihood of observing such a process. We describe the Advanced
Composition Explorer (ACE) satellite which we use to monitor the solar
wind at times of interest.

In Section \ref{secmethod} of the paper we detail the steps employed
to identify observations possibly affected by SWCX emission and in
Section \ref{secres} we present the results from the use of this
method for an initial small sample of approximately 170
observations. We conclude by discussing our plans for the application
of this method to incorporate data from the entire publicly available
XSA from launch until the commencement of this second phase. As this
time period will incorporate more than seven years of XMM-Newton
operations, we hope to detect any relationship between our results and
the solar cycle.

\subsection{Solar wind charge exchange}\label{secswcx}
SWCX occurs when an ion in the solar wind interacts with a neutral
atom and gains an electron in an excited state
(Eq. \ref{eqnswcxgive}). If the ion is in a sufficiently high charge
state in the subsequent relaxation of the ion, an X-ray or UV photon
is released (Eq. \ref{eqnswcx}).

\begin{equation}
A{^q}{^+} + B \to  A^{(q-1)^{\ast}} + B{^+}
\label{eqnswcxgive}
\end{equation}
\begin{equation}
A^{(q-1)^{\ast}} \to A^{(q-1)} + h\nu
\label{eqnswcx}
\end{equation}

A is the solar wind ion, for example oxygen, and B the neutral atom,
commonly hydrogen or helium in the heliosphere or just hydrogen in the
case of geocoronal neutrals in the Earth's magnetosheath.

A SWCX spectrum is characterised by emission lines corresponding to
the ion species present in the solar wind. Although the solar wind is
approximately 99\% protons, electrons and $\alpha$-particles, the
remainder of the wind consists of heavier elements such as C, O, Mg,
Si, Fe and Ni. The heavier elements found in the solar wind are often
of high charge state due to the high temperatures of the solar
corona. The relative abundances of the elements are dependent on the
conditions of the Sun when the solar wind leaves the solar corona. In
this project we concentrate our search on oxygen emission lines as
these are typically the strongest SWCX lines in previously published
SWCX-enhanced XMM-Newton observations, e.g. \citet{snowden2004}.

SWCX occurs at several locations in the solar system including
cometary tails \citep{greenwood2000, cravens1997}, at the heliospheric
boundary \citep{cravens2000, koutroumpa2007}, planetary atmospheres
\citep{Dennerl2002, branduardiraymont2007} and from the Earth's
magnetosheath or interplanetary neutrals \citep{cravens2001,
  wargelin2004, snowden2004, fujimoto2007}. We concentrate on charge
exchange between the solar wind and geocoronals in the Earth's
magnetosheath and contributions within the heliosphere in the vicinity
of the Earth, as this X-ray emission is the most likely to affect the
observations of XMM-Newton in terms of a short, temporally varying
background source. Local x-ray intensities vary on much shorter
timescales (on the order of hours) than those of the heliospheric-ISM
boundary contributions, which vary on much longer timescales and with
less variation in intensity \citet{robertson2003b}. The main variation
seen in the heliospheric contribution occurs in conjunction with
changes in the line of sight. As the solar wind travels at
approximately $\frac{1}{4}$ AU per day, the heliospheric emission is
the result of the average conditions of the solar wind over several
months.

\citet{robertson2003a}, \citet{robertson2003b} and
\citet{robertson2006} have modelled the expected X-ray emission from
charge exchange with geocoronal neutrals, using \citet{hodges1994}
model of the exosphere to account for the densities of the geocoronal
neutrals with height for differing solar wind conditions, mapping the
area of strongest X-ray emission on the sunward side of the
magnetosheath (at an intensity of approximately 8 keV
cm$^{-2}$s$^{-1}$ sr$^{-1}$ for nominal solar wind conditions, rising
to 200 keV cm$^{-2}$ s$^{-1}$ sr$^{-1}$ for simulated solar storm
conditions). The solar wind speed can vary dramtically depending on
storm or slow wind conditions. A slow solar wind of approximately 300
km s$^{-1}$ and average solar wind densities results in a nominal
solar proton flux of approximately 2$\times$10$^{8}$ cm$^{-2}$
s$^{-1}$. The exact distances from the Earth to the magnetopause (the
boundary layer between the solar wind plasma and the plasma of the
Earth's magnetosphere) and bow shock (which together define the
magnetosheath region) depend on the strength of the solar wind at any
given time \citep{khan1999} but are on average at approximately 9 and
15 R$_{E}$ respectively.

SWCX was first suggested as an explanation for the temporal variations
of a few hours to a few days length in the diffuse emission data of
the ROSAT All-Sky Survey \citep[long-term
enhancements,][]{snowden1995} by \citet{cox1998} and
\citet{cravens2000}. SWCX has been seen by XMM-Newton, Chandra and
Suzaku, and has been attributed to geocoronal emission
(e.g. \citet{fujimoto2007}). Multiple observations with the same
pointing direction of the Hubble Deep Field North region taken with
XMM-Newton have shown temporal variations in count rates and line
emission that has been correlated to enhancements in the level of the
solar wind flux and the solar storm of May 2001
\citep{snowden2004}. Also, depending on viewing geometry, it would be
possible to look along the front of a Coronal Mass Ejection (CME)
density enhancement, which could cause short term and extreme
variations in the level of X-ray flux observed \citep{collier2005}.

\subsection{The XMM-Newton observatory and viewing geometry}\label{secxmm}
The XMM-Newton observatory \citep{jansen} has been used extensively in
the study of extended and diffuse X-ray sources. For such studies, a
good understanding of the X-ray background is required, including the
SWCX, which may affect a user's observation.

Due to the supposed short term variations in the SWCX at the Earth's
magnetosheath an observation may be partly or wholly affected by
enhancements at energies characteristic of the SWCX line emission. The
effect is also of interest in its own right as a monitor of the ionic
composition of the solar wind and the interaction between the Earth's
magnetosheath and the bombardment of the solar wind.

\begin{figure}[]
\vspace{1cm}
\includegraphics[width=0.45\textwidth]{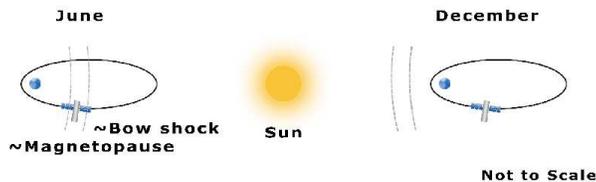}
\caption{A schematic illustratration of the orbit of XMM-Newton in
  relation to the magnetopause and bow shock (dotted lines), which
  define the boundaries of the magnetosheath, at different times
  during the year. The left image shows the highly elliptical orbit of
  XMM-Newton during mid summer (northern hemisphere), and the
  right image shows the winter configuration. The Sun has been added
  to show the direction of the Sun and therefore incoming solar wind.}
\label{figxmmorbit}
\end{figure}

XMM-Newton is not able to study the SWCX at all times due to the
configuration of it's highly elongated elliptical orbit. Optimum
viewing of the magnetosheath and areas of the expected highest X-ray
emission within the magnetosheath \citep{robertson2003b} is expected
to occur at only certain points during the 48 hour orbit at certain
times of the year, see Figure \ref{figxmmorbit}. The situation is
further complicated by restrictions on the viewing angle, which is
strongly constrained by the fixed solar panels and those imposed to
protect the instruments from directly viewing the Sun and the X-ray
and optically bright Earth and Moon, and the non-operational period
during the telescope's passage through perigee as it passes through
the radiation belts. This results in the observatory being able to
view the area of maximum SWCX emission, i.e. the nose of the
magnetosheath, only at certain times of the year. XMM-Newton is
therefore more likely to experience SWCX-emission enhancement during
the summer months (northern hemisphere) as this is when the telescope
is able to be in a position to look through the nose and bright flanks
of the magnetosheath. However, our goal is to study all observations
in the XMM-Newton archive assuming they were conducted in an
appropriate observational mode. Coupled with the information regarding
the solar wind flux we will, with this project, gain understanding
regarding the likelihood and level of SWCX-contamination an XMM-Newton
observation may experience.

\subsection{The Advanced Composition Explorer (ACE)}\label{secace}
The Advance Composition Explorer (ACE) was launched in 1997, into
orbit at the Earth-Sun Langrangian L1 point about 1.5 million miles
from Earth. The instrument package on ACE was designed to study solar,
galactic and extragalactic energetic particles (see:
\textit{http://www.srl.caltech.edu/ACE/ace\_mission.html}).

In this project we use ACE to provide additional supporting evidence
that a SWCX-enhancement may have been seen in an XMM-Newton
observation. We use hourly average data (\textit{Level 2}) from the
Solar Wind Electron, Proton, and Alpha Monitor (SWEPAM) instrument on
board ACE. Correlations of increased emission in the SWCX lines with
increases in the solar wind proton flux lend credence that a
particular observation may have experienced a
SWCX-enhancement. However the exact scenario maybe complicated by the
orientation of the interplanetary magnetic field, adjusting the delay
time between ACE and Earth \citep{weimer2003}, so conclusions drawn
from correlations between increased X-ray and solar proton flux should
be treated with caution. We use data from ACE as a general monitor of
the condition of the solar wind flux. However, the delay between ACE
and the Earth is approximately one hour.

\section{Method}\label{secmethod}
In this section we describe the steps taken in the initial data
reduction and then continue to detail the method used to test for the
presence of SWCX contamination.

The objective when designing the method was to find the key indicators
for SWCX-emission, namely short term variability in line emission from
the solar wind ion species. Short term variations in the diffuse
background indicate a local source, and the presence of line emission
at certain energies expected from the solar wind is a diagnostic of
SWCX-emission. We grade the indicators and develop a procedure to
automatically flag observations that may have experienced
SWCX-enhancement throughout their exposure. In this paper we consider
single XMM-Newton pointings, but the future paper will explore
multiple pointings of the same target to investigate variations
between pointings.

\begin{table*}
  \caption[]{{\small Summary of all the observations from the XSA
        that were used in this analysis. We list the observation identifier, the date of the
        observation and the exposure of the observation}} {\tiny{
\begin{tabular}{|ccc|ccc|ccc|}
  \hline
  Obs. ID. & Date & Exposure (s) &  Obs. ID. & Date & Exposure & Obs. ID. & Date & Exposure (s) \\ 
  \hline
  0099760201001 & 2000-03-22 & 49332.0 & 0113891001002 & 2000-04-06 & 11605.6 & 0113891101002 & 2000-04-07 & 16888.5 \\
  0125300101012 & 2000-05-28 & 32199.8 & 0124711401002 & 2000-05-29 & 17663.9 & 0125920201002 & 2000-06-05 & 23447.8 \\
  0126700401008 & 2000-06-15 & 7344.45 & 0124710401002 & 2000-06-23 & 8302.49 & 0124711101002 & 2000-06-24 & 28264.6 \\
  0127921001001 & 2000-07-21 & 53637.1 & 0127921201001 & 2000-07-23 & 18456.8 & 0127921101001 & 2000-07-23 & 7264.02 \\
  0127720201001 & 2000-07-25 & 22364.9 & 0112580601001 & 2000-07-26 & 33567.5 & 0111210201001 & 2000-07-28 & 6135.45 \\
  0101440101001 & 2000-09-05 & 49196.7 & 0101440701002 & 2000-09-05 & 29643.4 & 0105460301002 & 2000-09-07 & 18773.5 \\
  0104460401002 & 2000-10-12 & 11898.8 & 0105260201001 & 2000-10-13 & 17851.8 & 0123720301003 & 2000-10-27 & 56029.2 \\
  0111400101001 & 2000-11-03 & 50115.7 & 0106460101001 & 2000-11-06 & 54873.2 & 0101440401001 & 2000-11-07 & 45359.8 \\
  0101040301001 & 2000-11-28 & 37070.4 & 0093552701001 & 2001-01-28 & 24138.1 & 0086360401005 & 2001-03-12 & 44171.7 \\
  0064940101001 & 2001-03-24 & 36378.6 & 0083250101004 & 2001-04-09 & 19805.2 & 0112870201002 & 2001-04-17 & 15116.5 \\
  0111550401002 & 2001-06-01 & 91601.5 & 0059140201001 & 2001-06-17 & 12191.9 & 0051760101001 & 2001-06-17 & 7053.58 \\
  0051760201001 & 2001-06-18 & 15740.6 & 0109470201001 & 2001-07-11 & 12481.7 & 0137750101001 & 2001-07-29 & 17501.8 \\
  0109100201002 & 2001-08-17 & 9935.07 & 0000110101001 & 2001-08-19 & 23847.7 & 0089210601001 & 2001-10-11 & 54619.6 \\
  0089210701001 & 2001-10-11 & 13630.0 & 0089210701002 & 2001-10-11 & 22315.9 & 0085150201002 & 2001-10-21 & 37300.7 \\
  0085150301003 & 2001-10-21 & 31526.0 & 0022740301001 & 2001-11-04 & 35347.6 & 0028740101001 & 2001-11-15 & 28144.9 \\
  0089210101001 & 2001-12-01 & 8354.39 & 0084230201001 & 2001-12-14 & 28220.5 & 0083000301001 & 2001-12-15 & 26436.8 \\
  0113050501001 & 2001-12-23 & 25154.6 & 0084030101001 & 2001-12-28 & 42853.5 & 0001930301001 & 2001-12-28 & 24582.5 \\
  0106860201001 & 2001-12-28 & 15353.4 & 0002740301001 & 2001-12-29 & 6847.05 & 0093200101002 & 2001-12-29 & 39631.2 \\
  0084140501002 & 2002-02-04 & 17436.3 & 0065820101001 & 2002-02-27 & 49238.9 & 0011420201001 & 2002-03-17 & 40687.2 \\
  0110661601002 & 2002-03-19 & 7267.89 & 0058940101002 & 2002-03-20 & 27788.2 & 0136000101002 & 2002-04-17 & 17691.4 \\
  0033540601001 & 2002-05-11 & 5224.01 & 0059140901001 & 2002-05-22 & 15889.5 & 0082140301001 & 2002-05-22 & 32887.9 \\
  0094360601001 & 2002-05-23 & 8236.27 & 0041750101001 & 2002-06-15 & 51641.7 & 0134521601011 & 2002-06-18 & 11043.9 \\
  0134521601003 & 2002-06-18 & 24111.6 & 0031740101001 & 2002-06-19 & 35180.5 & 0135745801001 & 2002-09-29 & 13120.2 \\
  0112980401001 & 2002-09-30 & 16528.9 & 0148850201001 & 2002-10-01 & 23730.5 & 0089370501001 & 2002-10-01 & 22454.5 \\
  0147511101001 & 2002-10-23 & 97292.4 & 0150050101002 & 2002-11-26 & 23751.9 & 0147511801001 & 2002-12-06 & 90684.3 \\
  0112260801001 & 2002-12-07 & 13190.8 & 0112230201001 & 2002-12-18 & 25541.5 & 0150480501001 & 2002-12-22 & 21904.7 \\
  0070340101002 & 2003-02-02 & 17023.8 & 0152680801001 & 2003-02-26 & 16371.9 & 0157360601003 & 2003-02-26 & 6618.36 \\
  0146390201001 & 2003-03-29 & 25388.8 & 0153220601001 & 2003-05-28 & 10902.0 & 0150680101001 & 2003-07-26 & 42666.2 \\
  0152131201002 & 2003-08-16 & 9502.71 & 0049540301001 & 2003-08-17 & 25734.6 & 0049540401001 & 2003-08-19 & 24037.2 \\
  0149630301001 & 2003-09-16 & 19768.3 & 0017740401007 & 2003-10-05 & 27582.9 & 0017740201007 & 2003-10-09 & 17879.9 \\
  0017740601007 & 2003-10-11 & 22288.5 & 0017740701007 & 2003-10-13 & 21736.3 & 0147760101001 & 2003-10-14 & 36842.6 \\
  0017740501007 & 2003-10-20 & 26551.5 & 0148742501001 & 2003-10-28 & 7917.18 & 0162160201002 & 2003-11-24 & 14003.8 \\
  0162160401002 & 2003-12-06 & 10509.2 & 0162160601002 & 2003-12-14 & 12360.0 & 0205260201009 & 2004-01-16 & 13440.0 \\
  0201230201001 & 2004-01-21 & 18093.1 & 0200240301001 & 2004-04-02 & 15255.5 & 0164560701001 & 2004-07-23 & 31608.0 \\
  0164560901001 & 2004-09-12 & 58437.8 & 0205970201001 & 2004-09-20 & 48145.6 & 0206060201001 & 2004-11-03 & 31485.3 \\
  0202610801001 & 2004-11-09 & 7179.57 & 0201440101001 & 2004-11-09 & 10311.2 & 0202610801002 & 2004-11-09 & 17784.4 \\
  0201903101001 & 2004-11-10 & 27833.5 & 0201900501001 & 2004-11-12 & 27163.9 & 0203360901001 & 2004-11-20 & 27258.5 \\
  0203240201017 & 2004-12-04 & 18804.6 & 0203240201001 & 2004-12-05 & 19919.2 & 0203360101001 & 2004-12-11 & 30644.7 \\
  0200630201001 & 2004-12-20 & 44644.7 & 0202370301001 & 2005-01-08 & 25837.4 & 0202130301001 & 2005-01-15 & 37385.1 \\
  0212480701002 & 2005-06-05 & 19008.7 & 0211280101002 & 2005-06-12 & 42621.5 & 0304531501006 & 2005-06-22 & 15215.0 \\
  0305920601001 & 2005-06-23 & 15224.2 & 0304050101001 & 2005-06-30 & 69689.5 & 0212480801001 & 2005-07-01 & 48830.5 \\
  0302420101001 & 2005-07-08 & 85607.2 & 0304531701002 & 2005-07-10 & 7502.96 & 0164571401001 & 2005-08-21 & 56486.1 \\
  0300800101002 & 2005-08-31 & 45875.7 & 0300540101001 & 2005-09-14 & 23398.9 & 0303100101001 & 2005-09-26 & 52629.3 \\
  0300600401001 & 2005-10-08 & 28648.7 & 0305560101001 & 2005-10-21 & 27595.6 & 0302310501002 & 2005-10-23 & 23162.3 \\
  0159760301001 & 2005-11-01 & 37543.4 & 0302351601007 & 2005-11-27 & 43034.0 & 0302310301002 & 2005-12-18 & 43998.8 \\
  0303560201001 & 2005-12-19 & 6516.42 & & & & & & \\ 
  \hline
\end{tabular}
\label{tabobsused}
}}
\end{table*}

\subsection{Data reduction}\label{secdatared}
The data used in this analysis were taken from the XSA, which can be
found at: \indent \textit{http://xmm.vilspa.esa.es/external/}\newline
\indent \indent \textit{xmm\_data\_acc/xsa/index.shtml}.

We used publicly available software in this project, accessible
through the web pages of the XMM-Newton Background Working Group
(BGWG):\\ \indent \textit{
http://xmm.vilspa.esa.es/external/xmm\_sw\_cal/ \\ \indent \indent
background/index.shtml}

For each observation an Original Data File (ODF) was downloaded from
the XSA. The ODF files were processed using the XMM-Newton Extended
Source Analysis Software (ESAS) software package \citep{kuntz2008,
  snowden2008}, found on the BGWG pages as mentioned above.

Table \ref{tabobsused} lists the observations used in this preliminary
analysis. We included several observations used by \citet{kuntz2008}
and \citet{snowden2004} with the aim to use these observations as
control subjects for our method. The remainder of the observations
were taken from a range of times of year and times in the mission,
representive as much as possible of a random selection of observations
throughout the XMM-Newton mission.

To undertake our analysis we created lightcurves from event files for
each XMM-Newton observation. To do this we filtered the data files for
soft proton and particle background contamination using ESAS and other
software before commencing with the creation of the lightcurves.

\begin{figure}
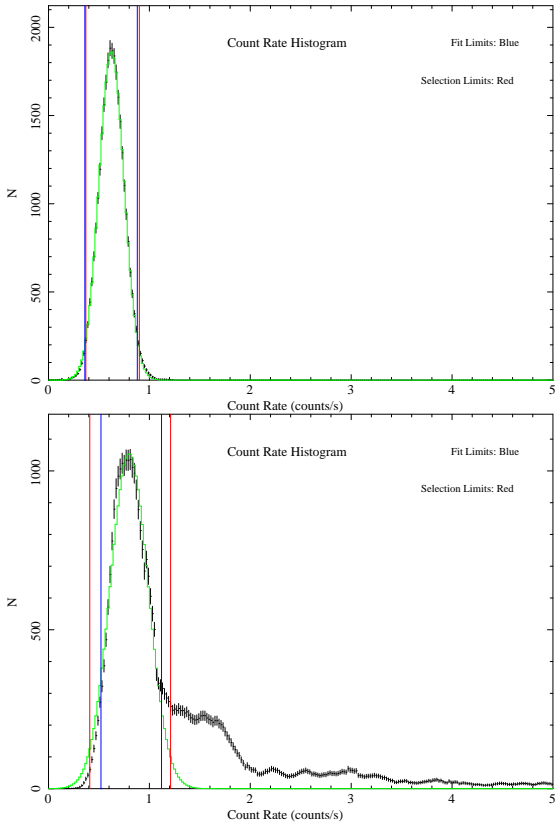

      \includegraphics[width=.3\textwidth, angle=270]{new_examplecr_good_crop.ps}
      \includegraphics[width=.3\textwidth, angle=270]{new_examplecr_bad_crop.ps}
      \caption{Example count rate histograms from the ESAS software
        where a Gaussian shaped curve is fitted to the histogram. The
        upper panel shows a count rate histogram where the data and
        the Gaussian curve are almost coincident and little or no
        flaring has occurred, whereas the lower panel shows flaring
        characteristics.}
      \label{figexphist}
\end{figure}

The ESAS software is currently only available for the MOS detectors
\citep{turner} and therefore the analysis detailed here only concerns
the MOS instruments. However, improvements to include the PN
\citep{struder} camera are underway and are expected in an imminent
release of the ESAS software. In our future paper we aim to
incorporate PN data into our analysis. Cleaned and calibrated event
files were created from the ODF when using the ESAS tool
\textit{mos-filter}. The task \textit{mos-filter} runs several
XMM-Newton Science Analysis Software (SAS) tasks, including
\textit{emchain} for the basic processing of the event files. It then
created two lightcurves; one in the field of view and one outside of
the field of view from the corners of the detectors for a band between
2.5\,keV and 12\,keV. Soft proton contamination will fall in the field
of view (with the exception of extremely large flares where some soft
proton contamination may be scattered into the out of field of view
regions), whereas the out of field of view data are only affected by
the particle-induced background. A count-rate histogram is then
created for the in field of view lightcurve, which should have a
roughly Gaussian profile if the observation is unaffected by
soft-proton contamination (see the example in Figure
\ref{figexphist}). Any large bumps or deviations from the roughly
Gaussian shape should be treated with caution as these indicate high
levels of contamination. A Gaussian is fitted to the peak of the
distribution and using a threshold of ${\pm}$1.5$\sigma$ this fit is
used to create good-time-interval (GTI) periods for the data. The blue
vertical lines of Figure \ref{figexphist} show the range used for the
Gaussian fit and the red vertical lines indicate the limits taken for
the GTI periods. Further SAS tasks are used to filter the files using
the GTI periods to create a cleaned event file. An important point to
note is that there may still be residual soft-proton contamination
after undergoing these procedures \citep{deluca2004}.

For each event file we removed sources in the field of view by
utilising the source list from the 2XMM catalogue processing (see
\newline (\textit{http://xmm.esa.int/external/xmm\_user\_support/}
\newline \indent \textit{documentation/uhb/node141.html}) to create a
region file and then extracting the sources from the event file with
\textit{evselect}. We initially used a source extraction radius of 35
arcseconds, extending this to remove bright or extended
sources if needed.

Checks were made of images created for each cleaned event file to
ensure that residual sources were not present. If residual sources
remained the extraction radius was increased iteratively until an
effectively source-free event file was obtained, although observations
with exceptionally bright sources were removed from the sample. The
remaining useful observations are shown in Table
\ref{tabobsused}. After this procedure any residual emission from
point sources should be negligible. Software for the automated removal
of sources will be developed for part II of these articles. We
considered observations that use full-frame mode and rejected all
other modes for this preliminary study.

We then created spectra and the associated response matrices and
auxiliary files for those observations which we wished to study
further. We used the ESAS tasks \textit{mos-spectra} and
\textit{mos-back} to create the spectra, background spectra and
instrument response and auxiliary files. We used a circular spectrum
extraction radius of 17000 detector units to incorporate the in field
of view region (pixel size of 0.05 arcseconds, 14.2 arcminute radius),
centred on \textit{DETX, DETY} coordinates of (0, 0).

\subsection{Examination for SWCX emission}\label{secdatavar}
The first test was to grade the variation in a combined MOS1 and MOS2
(or with a single instrument if this was the only one available)
lightcurve created in a low-energy band covering a SWCX line compared
to the variation seen in a lightcurve representing the continuum. The
continuum lightcurve covered events in the band 2500\,eV to
5000\,eV. These points were chosen to be in regions where no emission
lines would be found. The line emission lightcurve covered events
within the 500\,eV to 700\,eV band, to concentrate on the OVII and
OVIII emission at 560\,eV and 650\,eV. For a typical observation, the
bands for both line and continuum lightcurves were created so that the
total number of counts in each lightcurve were similar. Each
lightcurve was created using events with the data flag FLAG==0 (those
events which are not from a bad pixel, or next to a bad pixel) and the
histogram created with a bin size of 1000\,s.

Lightcurves were rejected from the sample if they were less than
5000\,s long or if greater than 10\% of the bins had less than 20
counts, for either lightcurve.

The lightcurves were filtered using the previously determined GTI file
(as part of the ESAS output), and the counts in each bin adjusted for
any reduction in exposure of that bin. Bins that had undergone severe
GTI filtering where more than 40\% of the bin was removed were
excluded.  The Poisson error of each time bin and each lightcurve
was computed and adjusted for any reduction in exposure of that bin.

\begin{figure}[ht]
  \subfigure[Example scatterplot where no SWCX is expected, showing a
  good line fit to the scatterplot producing a low \redc\
  value]{\includegraphics[width=0.45\textwidth]{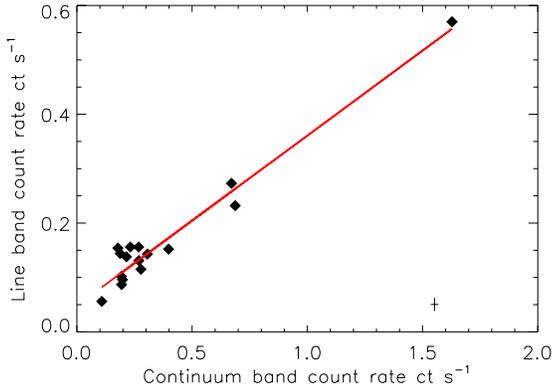}}
  \subfigure[Example scatterplot where SWCX has been observed, showing
  deviations from the line fit resulting in a high \redc\
  value]{\includegraphics[width=0.45\textwidth]{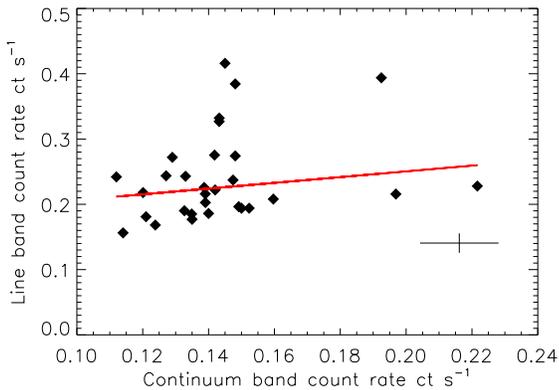}}
  \caption{Example scatter plots between the line and continuum bands
    for an observation that has experienced high levels of residual
    soft proton contamination, yet still shows a good linear fit
    between the line and continuum count rates (first panel) and an
    observation that shows SWCX-emission features (second
    panel). Representative average error bars are plotted in the
    bottom right-hand corner of each plot.}
\label{figexamplescat}
\end{figure}

For each set of lightcurves we created line-continuum scatter plots,
an example of which is shown in Figure \ref{figexamplescat}. If no
enhancement due to SWCX has occurred, we assumed correlation between
the line and continuum bands. Soft protons are spectrally variable in
intensity and shape but show no lines. If residual soft proton
contamination has occurred (which is most likely to be slowly varying
after the significant flare periods have been removed by GTI
filtering), the bands chosen for the line and continuum lightcurves
are sufficiently close that any spectral change in the continuum band
should be reflected in the line band, and therefore this assumption
holds. We looked at the scatter plot for several observations that
were expected to have experienced high levels of residual soft proton
contamination and indeed the straight line approximation was
valid. SWCX-enhanced observations were expected to show much less
correlation between the line and continuum bands. Bins that experience
SWCX-enhancement would show a deviation from the fit with increased
count rate levels in the line band.

A fit to each scatter plot is computed using the IDL procedure,
\textit{linfit}, which fits a linear model to a set of data by
minimising the $\chi^{2}$ statistic, see Equation \ref{eqnchisq}. We
use the continuum band count rate as the independent variable and the
values of the dependent variable are taken to be the count rate in the
line band.

\begin{equation}
\chi^{2} =  \sum_{i}^{}(D_{i} - E)^{2}/\sigma^{2}_{i}
\label{eqnchisq}
\end{equation}
\textit{D} are the points of the line-continuum scatter plot,
\textit{E} is the expected value, as found by the fit to the
line-continuum scatter plot and $\sigma$ the error on each point.\\

The procedure gives a $\chi^{2}$ for each fit to the scatter plot. We
divide this $\chi^{2}$ by the degrees of freedom of the fit to obtain
the reduced-$\chi^{2}$ for the fit, hereafter referred to as \redc.  A
high \redc\ indicates that a significant fraction of the points have a
large deviation from the best fit line found by the fit, and therefore
we expected that these cases would be more likely to show
SWCX-enhancement.

In addition we computed the $\chi^{2}$ values for each individual
lightcurve and calculate the ratio between the line and continuum band
$\chi^{2}$ values to add to our diagnostic (hereafter denoted as
\cratio).

As a measure of possible residual soft proton contamination, we
computed the ratio of in field of view to out of field of view counts
in a band between 7000\,eV and 12000\,eV, correcting for the area lost
due to the removal of point sources. To do this we used a script
\textit{fin\_over\_fout} which is publicly available through the BGWG
web-site (as previously mentioned) where a high ratio (hereafter
denoted as \fin) indicates a large residual soft proton contamination
component ($<$ 1.15 not contaminated, 1.15-1.3 slightly contaminated,
1.3-1.5 very contaminated, $>$ 1.5 extremely contaminated).  We also
considered the background time series files, as produced as part of
the 2XMM Serendipitous Source Catalogue \newline
((\textit{http://xmm.esa.int/external/xmm\_user\_support/} \newline
\indent \textit{documentation/uhb/node141.html})). \newline These
files contain lightcurves for the highest energy events (created for
full-field events with PI $>$ 14\,keV and with the selection
expression (PATTERN==0) \&\& \#XMMEA\_22 \&\& ((FLAG \& 0x762ba000) ==
0)). Large and/or frequent flaring periods in the time series would
suggest that the observation suffered from soft proton contamination
or that a residual component may still be present depending on the
effectiveness of the flare filtering procedure. Closer scrutiny of the
GTI filtering in such a case is required.

\section{Results}\label{secres}
In this section we present the general trends in the results from our
sample. We briefly discuss the results from the control subjects that
were added to our sample. These were observations with previously
detected SWCX-enhancement and have published results. We then discuss
several examples of previously unreported cases of SWCX-enhancement,
identified by the method described in this paper, and that produced
the highest \redc\ to the linear fit. The observations were ranked by
\redc\ and the top 30, for brevity, are summarised in Table
\ref{tabresults}. We have removed observations from the sample that
showed a significant deviation in the distribution of count rates from
the Gaussian that was determined and used by the ESAS software for
filtering purposes.

\begin{table*}
  \caption{Results for the observations which showed the highest 
    values of \redc. Each observation is given a case number (column one) and the observation identification number is shown in column two. 
    We give the date, the exposure time (Exp.), the degrees of 
    freedom ($\nu$) after any flare-filtering, the \redc\ value and then 
    state the ratio between the $\chi^{2}$ value of the line 
    lightcurve and that of the continuum lightcurve (\cratio). We also state 
    the average \fin\ (FF) ratio (and error on this value) between the MOS1 and MOS2. 
    The final column gives a comment about the 
    observation as described in the text.}
\begin{tabular}{|c|cccccccc|l|}
\hline
Case  & Observation  & Date & Exp. &$\nu$ &\redc&\cratio& FF    & Err. FF & Comment \\ 
      &              &      &  (ks)&    &linfit        &       & ratio & ratio    &  \\ 
\hline
1 & 0085150301 & 2001-10-21 & 31.96& 25 & 25.31 &  5.98 & 1.806 & 0.09 & Strong case SWCX \\
2 & 0093552701 & 2001-01-28 & 24.17& 16 & 22.97 &  6.09 & 1.486 & 0.09 & Weak case SWCX \\
3 & 0149630301 & 2003-09-16 & 19.77& 16 & 16.15 &  9.47 & 1.066 & 0.06 & Strong SWCX \\
4 & 0305920601 & 2005-06-23 & 15.24& 14 & 15.01 & 17.77 & 1.226 & 0.07 & Strong SWCX \\
5 & 0150680101 & 2003-07-26 & 42.67& 30 & 11.39 &  3.93 & 1.274 & 0.06 & Strong SWCX \\
6 & 0101040301 & 2000-11-28 & 37.21& 35 &  9.86 &  4.06 & 1.655 & 0.07 & Weak case SWCX \\
7 & 0111550401 & 2001-06-01 & 93.37& 83 &  8.53 &  5.36 & 1.339 & 0.04 & \citet{snowden2004}\\
8 & 0302310501 & 2005-10-23 & 23.16& 23 &  7.21 &  0.47 & 2.535 & 0.12 & Low \cratio \\ 
9 & 0127921101 & 2000-07-23 & 7.43 &  6 &  5.70 &  8.97 & 1.310 & 0.12 & \citet{kuntz2008} \\
10& 0070340501 & 2001-06-18 & 19.10&  8 &  5.43 &  1.85 & 1.760 & 0.14 & Weak case SWCX \\
11& 0202370301 & 2005-01-08 & 25.85& 14 &  4.85 &  1.19 & 1.202 & 0.07 & Low \cratio \\
12& 0127921001 & 2000-07-21 & 54.04& 53 &  4.73 &  1.93 & 1.660 & 0.07 & \citet{kuntz2008} \\
13& 0202610801 & 2004-11-09 & 17.90& 15 &  4.68 &  1.75 & 1.378 & 0.08 & No SWCX \\
14& 0136000101 & 2002-04-17 & 17.75& 17 &  4.12 &  1.56 & 1.570 & 0.09 & Strong case SWCX \\
15& 0150480501 & 2002-12-22 & 21.93& 11 &  3.82 &  0.75 & 1.467 & 0.11 & Low \cratio \\
16& 0112580601 & 2000-07-26 & 33.66& 32 &  3.80 &  1.95 & 2.171 & 0.11 & No SWCX \\
17& 0101440101 & 2000-09-05 & 49.22& 31 &  3.45 &  1.56 & 1.534 & 0.07 & Weak case SWCX \\
18& 0305560101 & 2005-10-21 & 23.01& 22 &  3.39 &  1.67 & 1.755 & 0.09 & No SWCX \\
19& 0001930301 & 2001-12-28 & 24.58& 18 &  3.21 &  1.37 & 2.010 & 0.10 & No SWCX \\
20& 0125920201 & 2000-06-05 & 23.45& 22 &  2.97 &  0.48 & 1.389 & 0.10 & Low \cratio \\
21& 0135745801 & 2002-09-29 & 13.16& 12 &  2.96 &  2.30 & 1.162 & 0.08 & No SWCX \\
22& 0089370501 & 2002-10-01 & 49.23& 22 &  2.90 &  1.40 & 1.143 & 0.06 & No SWCX \\
23& 0113050501 & 2001-12-23 & 25.64& 19 &  2.84 &  0.91 & 1.216 & 0.06 & Low \cratio\\
24& 0147511101 & 2002-10-23 & 97.33& 31 &  2.75 &  0.68 & 1.373 & 0.06 & Low \cratio\ \& bad flaring\\
25& 0164560701 & 2004-07-23 & 31.62& 20 &  2.74 &  1.53 & 1.520 & 0.07 & Weak case SWCX \\
26& 0099760201 & 2000-03-22 & 49.34& 35 &  2.69 &  0.74 & 1.436 & 0.06 & Low \cratio\\
27& 0085150201 & 2001-10-21 & 37.33& 21 &  2.43 &  0.25 & 1.834 & 0.09 & Low \cratio\\
28& 0106460101 & 2000-11-06 & 54.90& 43 &  2.41 &  1.06 & 1.308 & 0.06 & Weak case SWCX \\
29& 0000110101 & 2001-08-19 & 24.00&  9 &  2.31 &  0.83 & 1.355 & 0.09 & Low \cratio\\
30& 0205970201 & 2004-09-20 & 48.14& 35 &  2.24 &  0.42 & 1.466 & 0.06 & Low \cratio\ \& bad flaring\\
\hline
\end{tabular}
\label{tabresults}
\end{table*}

\subsection{General trends}\label{secgentrends}
\begin{figure}[h]
      \includegraphics[width=.45\textwidth]{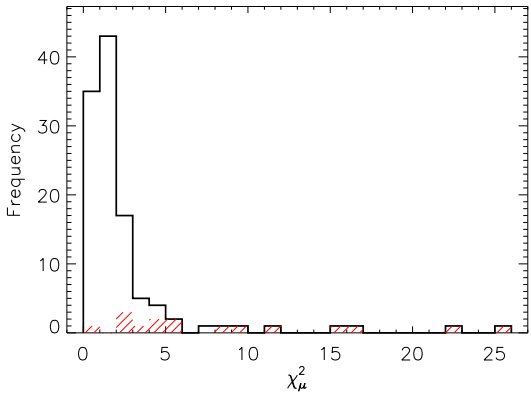}
      \caption{Histogram of \redc\ values.}
      \label{figredchist}
\end{figure}

In Figure \ref{figredchist} we present a histogram of the \redc\
values found for the whole data set. The modal value for the \redc\
value is between 1.5 and 2.0. We see that there are several cases with
extreme \redc\ values, showing the greatest disparity between line and
continuum count rates. We expect the SWCX cases to be found with the
highest \redc\ values. Shaded blocks in all figures in this section
indicate those observations with a suspected or previously published
SWCX enhancement, as will be discussed in the following sections.

\begin{figure}[h]
      \includegraphics[width=.45\textwidth]{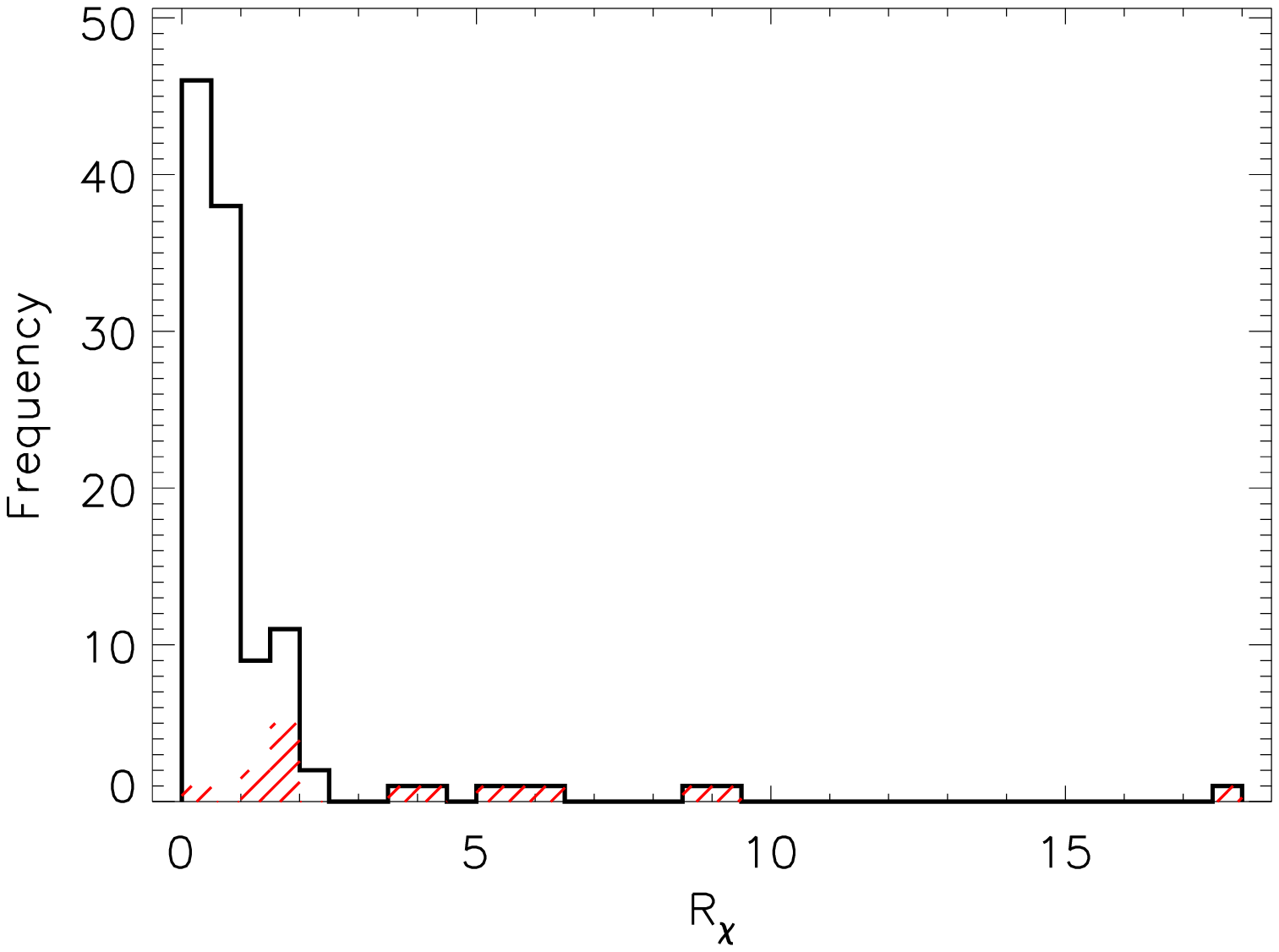}
      \caption{Histogram of \cratio.}
      \label{figcsrhist}
\end{figure}

In Figure \ref{figcsrhist} we present a histogram of the \cratio,
which is the ratio between the $\chi^{2}$ value for the line band
lightcurve to the $\chi^{2}$ value for the continuum band
lightcurve. SWCX-enhancements will have increased variation in the
line band compared to the continuum band, so a high \cratio\ is
expected for SWCX-affected observations. The shaded blocks show that
there are several observations where high variability is found along
with high \cratio\ values. High SWCX-enhanced observations are
expected for these cases.

\begin{figure}[h]
      \includegraphics[width=.45\textwidth]{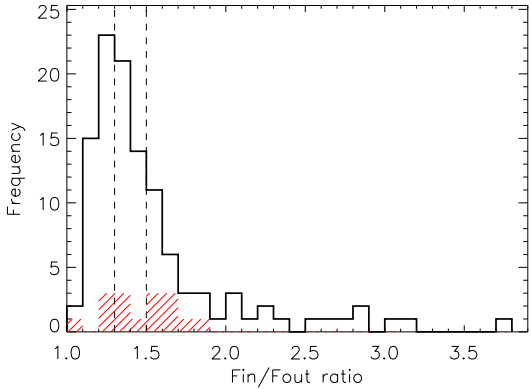}
      \caption{Histogram of \fin\ ratios. }
      \label{figffrhist}
\end{figure}

In Figure \ref{figffrhist} we present a histogram of the \fin\
ratios. This shows that twelve observations with a high \redc\ value
have been judged extremely contaminated by residual soft protons as
they show a \fin\ ratio $\geq$1.5. The two vertical dashed lines in
this figure indicate the \fin\ thresholds for classification as being
very contaminated (1.3-1.5) or extremely contaminated $>$1.5 by
residual soft protons.

\begin{figure}[h]
      \includegraphics[width=.45\textwidth]{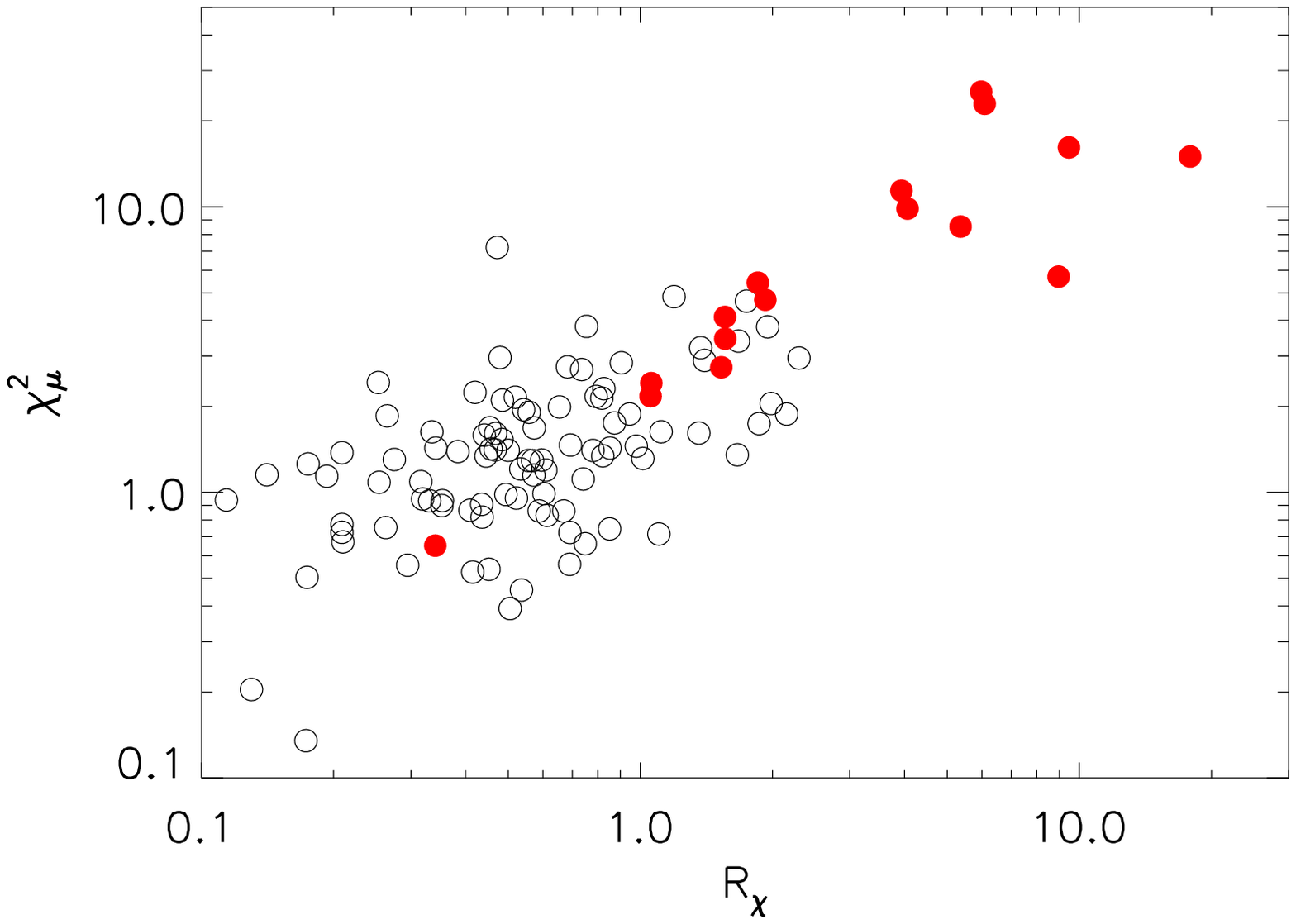}
      \caption{Scatter plot of \redc\ versus \cratio\ values. Filled
        circle are used for observations with a suspected or
        previously published SWCX enhancement.}
      \label{figscatredcvscr}
\end{figure}

\begin{figure}[h]
      \includegraphics[width=.45\textwidth]{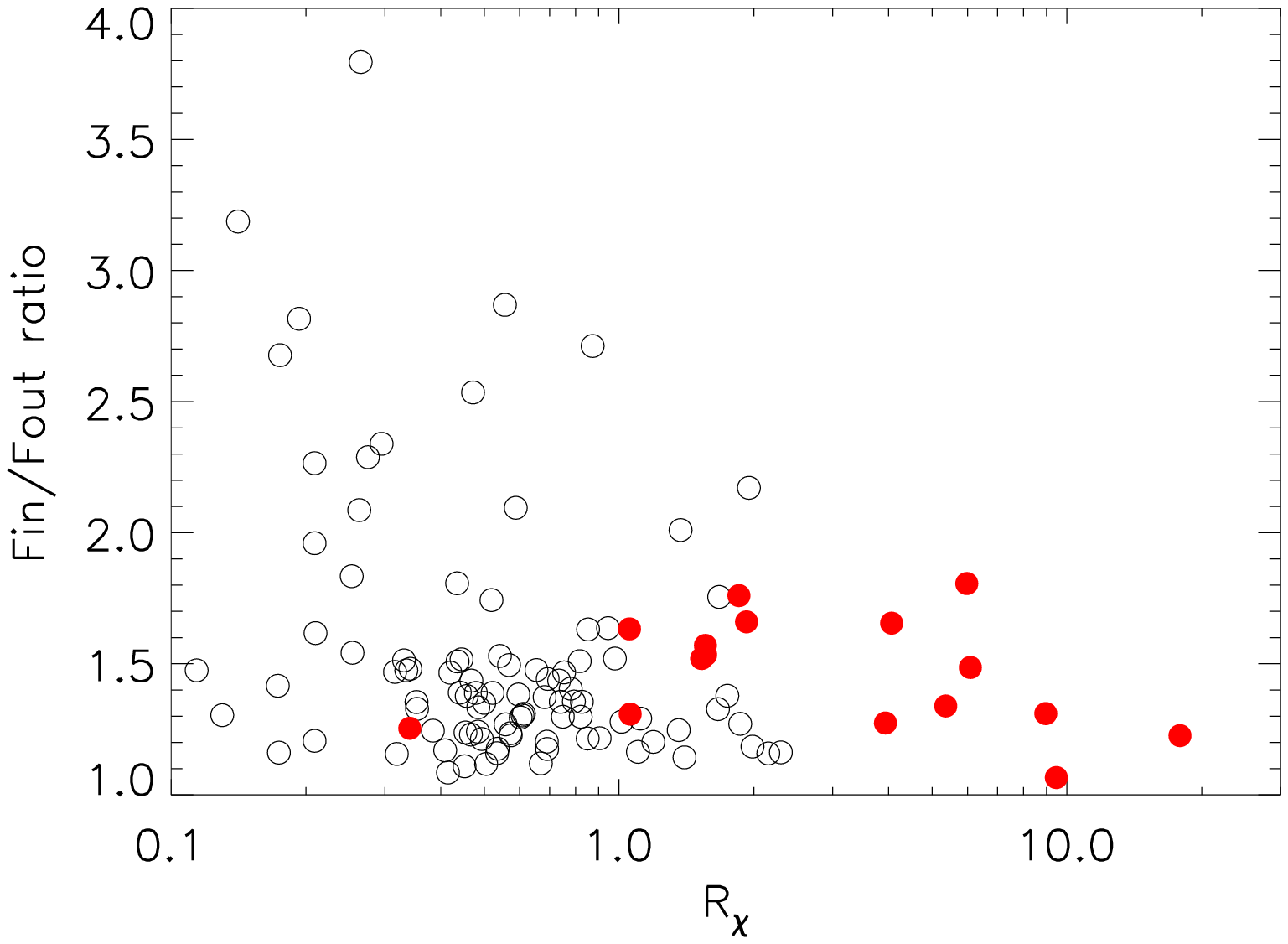}
      \caption{Scatter plot of \cratio\ versus \fin\ values. Filled
        circles are used for observations with a suspected or
        previously published SWCX enhancement.}
      \label{figscatcrvsfin}
\end{figure}

In Figure \ref{figscatredcvscr} we plot \redc\ values versus their
corresponding \cratio. In general, the highest \cratio\ are seen to
have the highest \redc, as expected for SWCX enhancements. The filled
circles, representing those observations with a suspected (this study)
or previously published SWCX enhancement, are found with the highest
\cratio, with one notable exception. These cases are discussed in
Section \ref{seccasesothers} and Section \ref{seccasesall}. The
exceptional case, where both the \cratio\ and \redc\ are low,
corresponds to a SWCX-enhanced control observation that is
unidentified by this method, and this case is discussed within Section
\ref{seccasesothers}. The distribution in this Figure suggests that
one may be able to start to define a threshold, between the parameters
\redc\ and \cratio, whereby a user would suspect that SWCX was
extremely likely to have occurred. This will be addressed further in
Paper II. In Figure \ref{figscatcrvsfin} we plot \cratio\ values
versus their corresponding \fin\ values. SWCX-enhanced observations
show a range of \fin\ values.

\subsection{Cases matching previously identified SWCX-enhanced observations}\label{seccasesothers}
\begin{figure*}
     \centering
     \subfigure[Case $\{$12$\}$, GWS1]{
          \includegraphics[width=.24\textwidth]{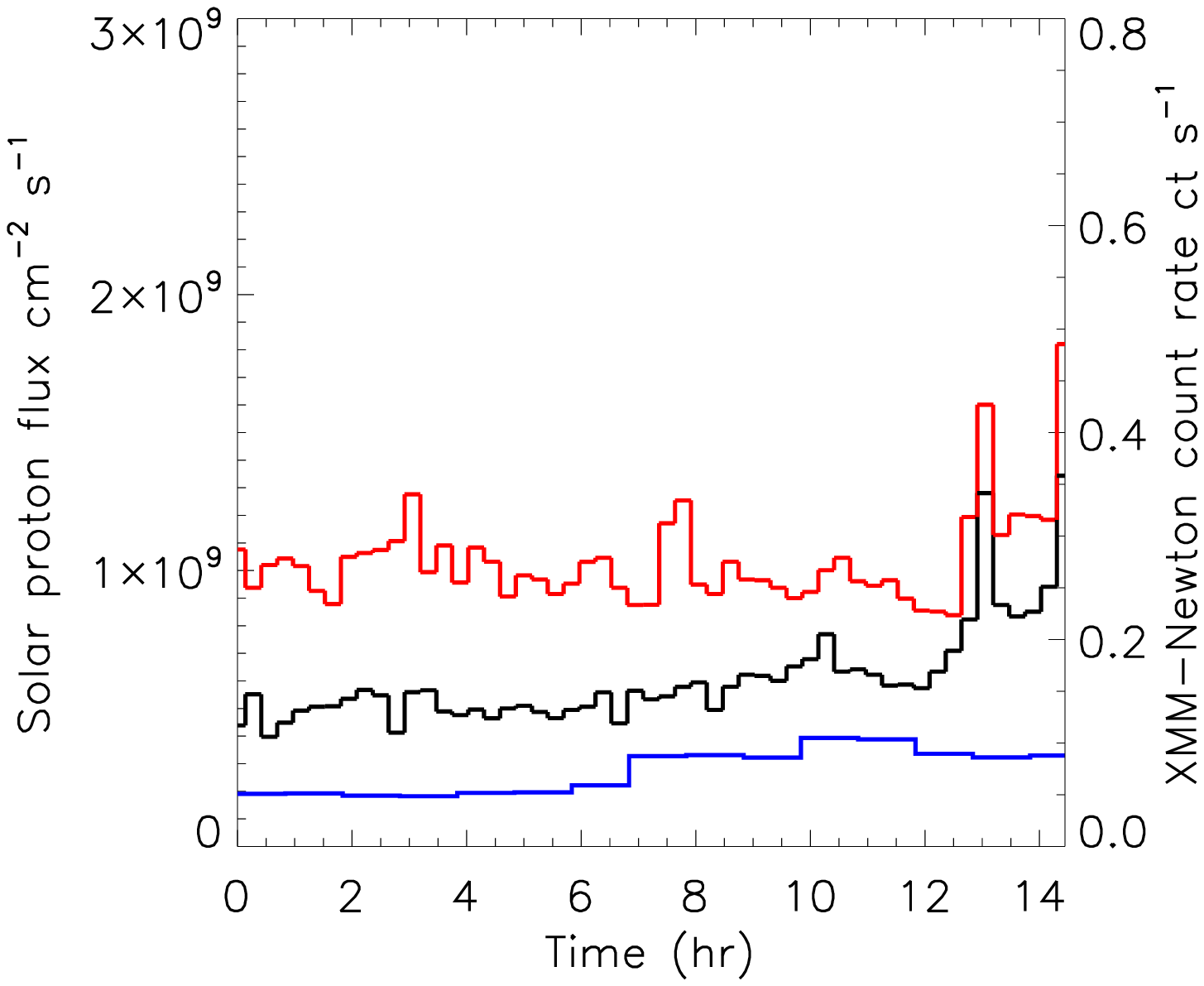}}
     \hspace{0.6cm}
     \subfigure[Case. $\{$9$\}$, GWS2]{
          \includegraphics[width=.24\textwidth]{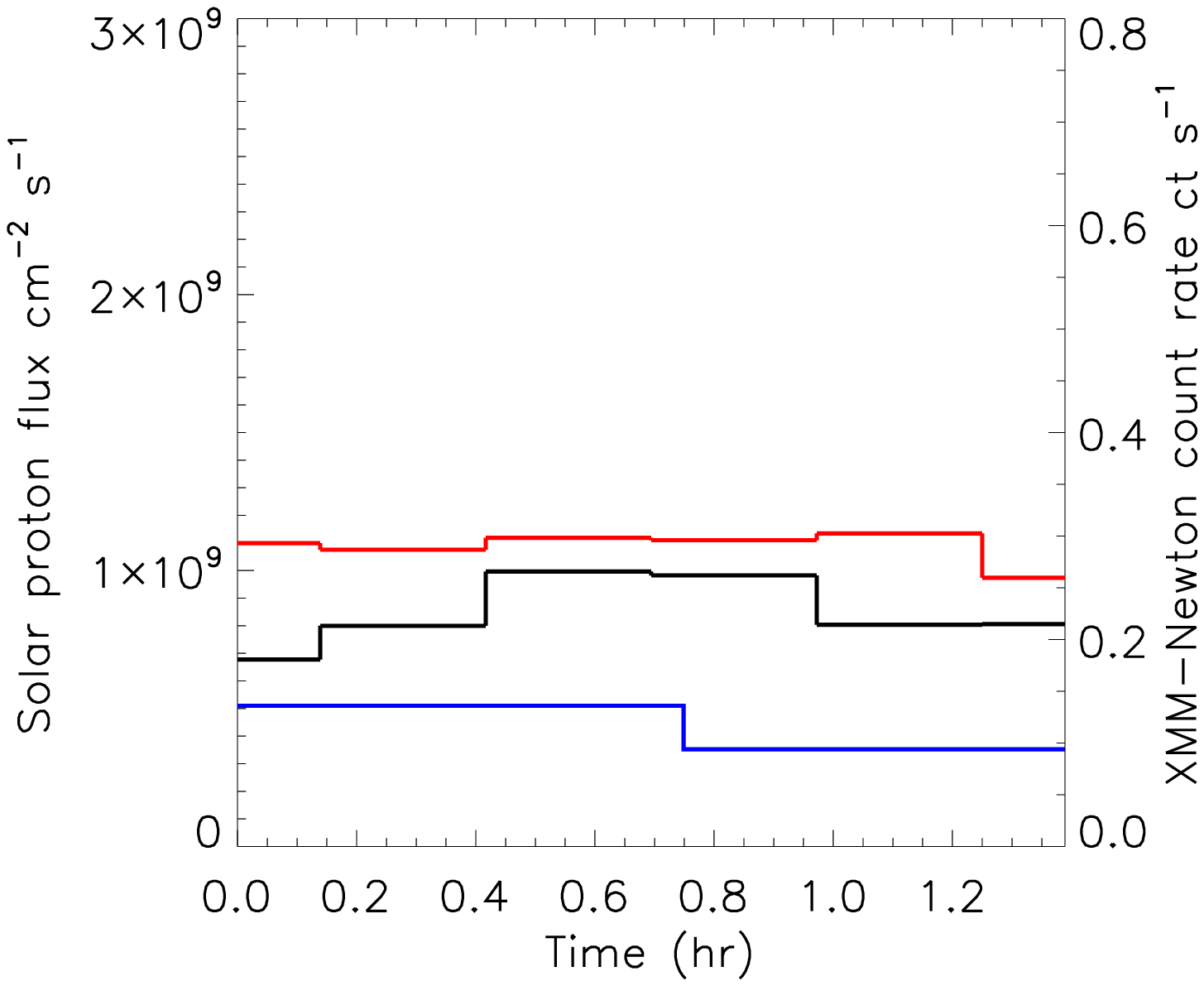}}
     \hspace{0.6cm}
     \subfigure[Case. $\{$63$\}$, GWS3]{
          \includegraphics[width=.24\textwidth]{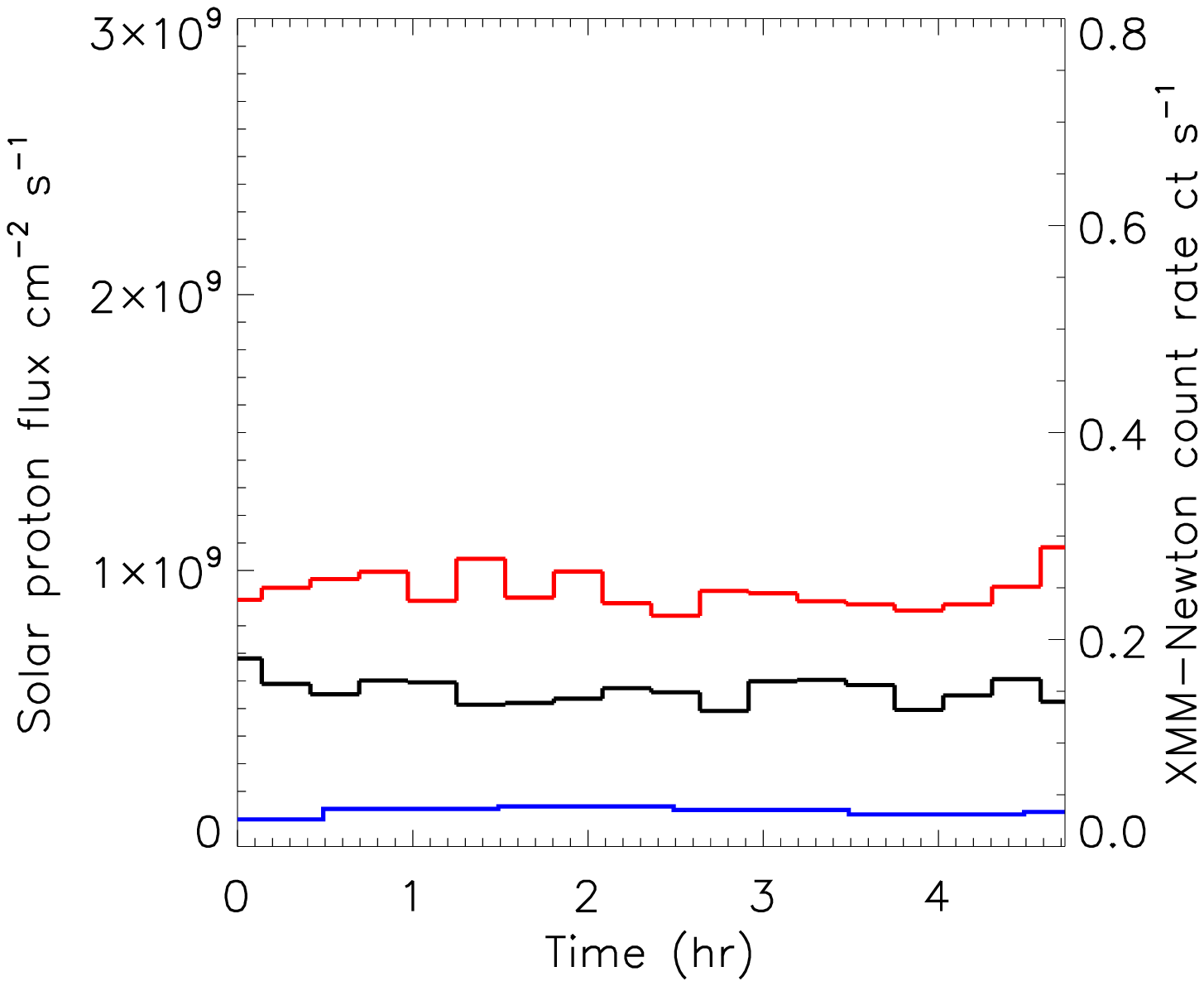}}
        \caption{Lightcurves for observations of the Groth-Westhall
          Strip. Blue represents the solar proton flux, black the line
          band count rate and red the continuum band count rate. These
          observations were studied spectrally by \citet{kuntz2008}
          and cases $\{$13$\}$ and $\{$10$\}$ were found to have
          SWCX-enhancement, whereas $\{$64$\}$ was devoid of SWCX
          features.}
     \label{figgwslcs}
\end{figure*}

\begin{figure*}
     \centering
     \subfigure[Case $\{$12$\}$, GWS1]{
          \includegraphics[width=.3\textwidth]{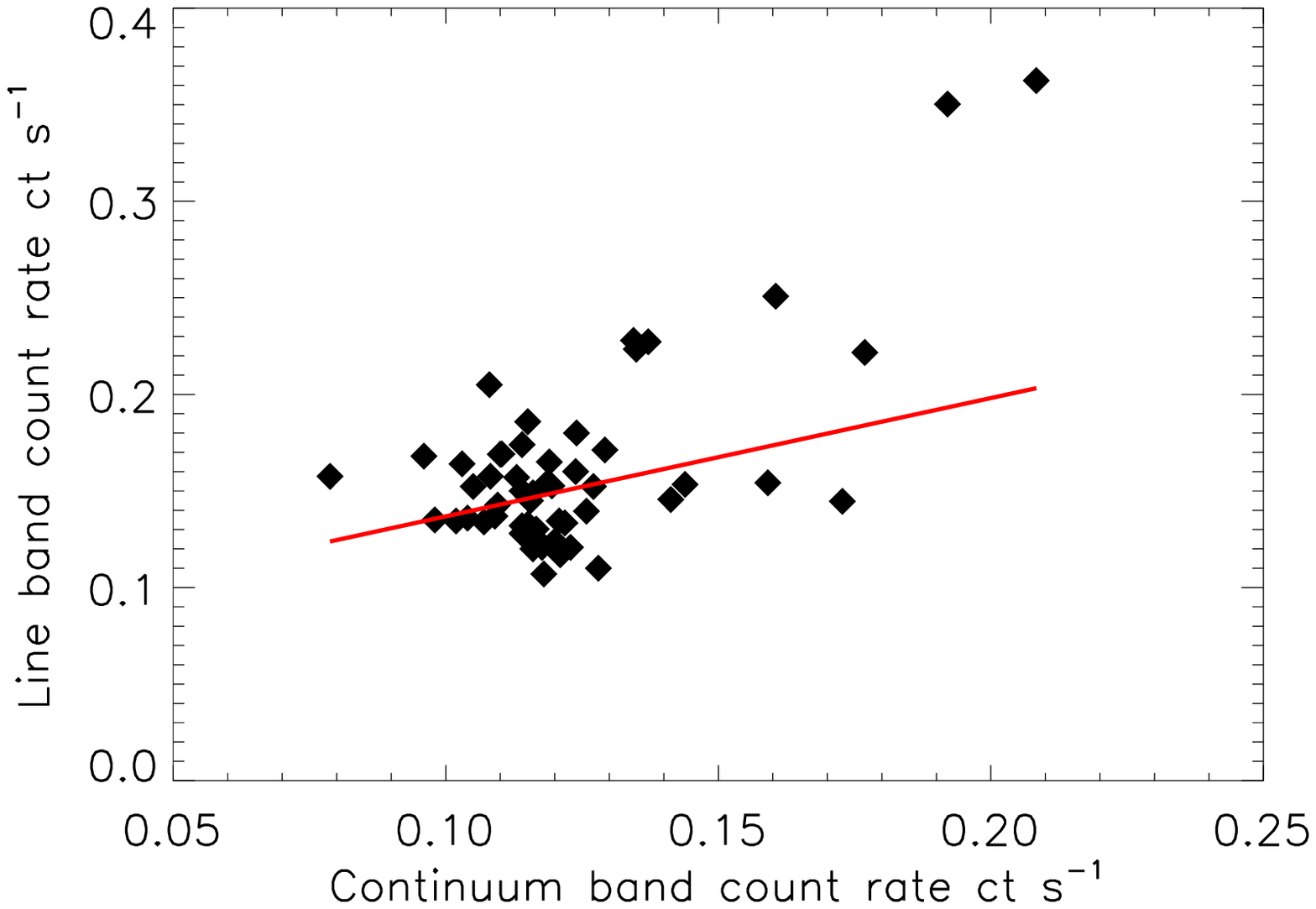}}
     \subfigure[Case $\{$9$\}$, GWS2]{
          \includegraphics[width=.3\textwidth]{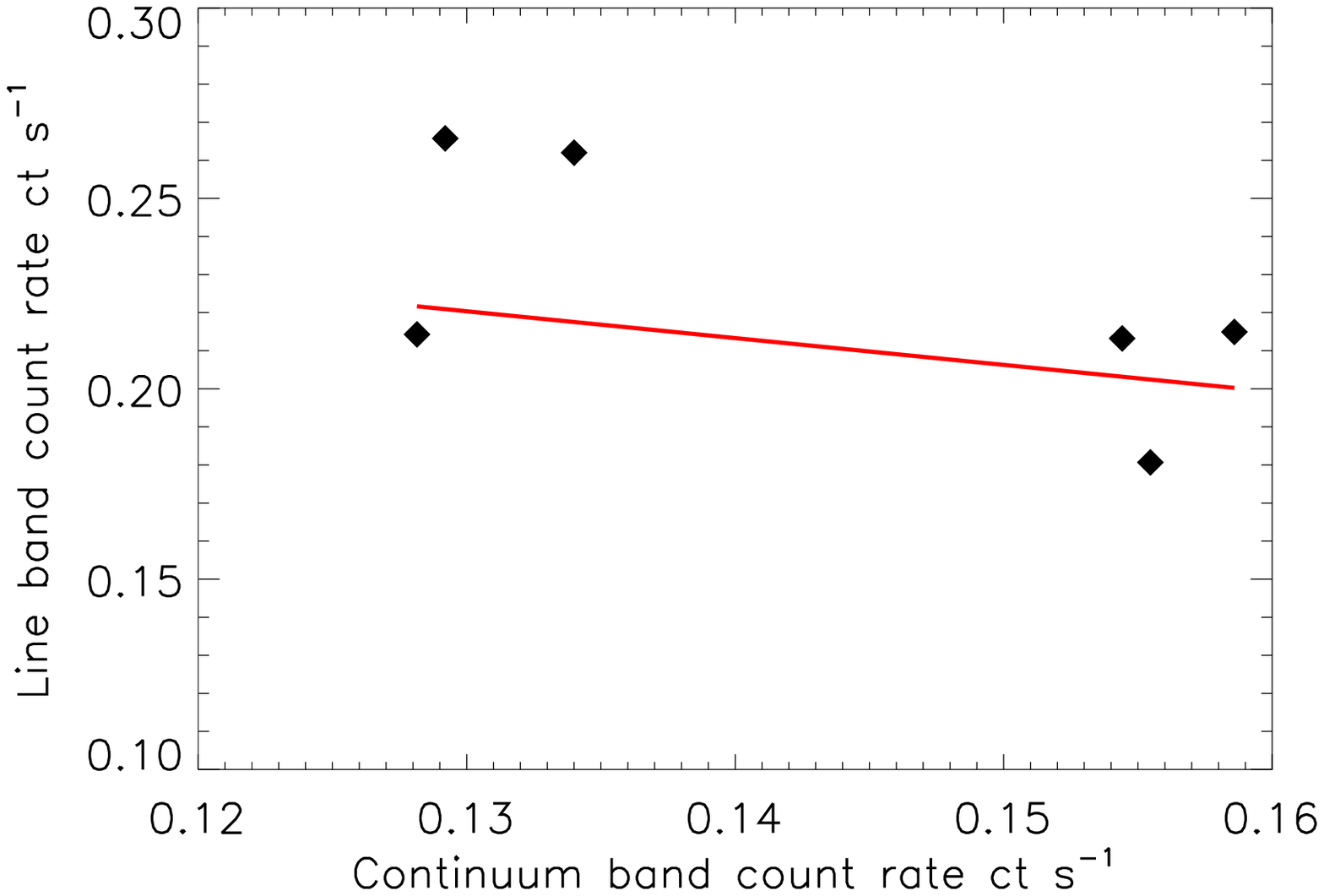}}
     \subfigure[Case $\{$63$\}$, GWS3]{
          \includegraphics[width=.3\textwidth]{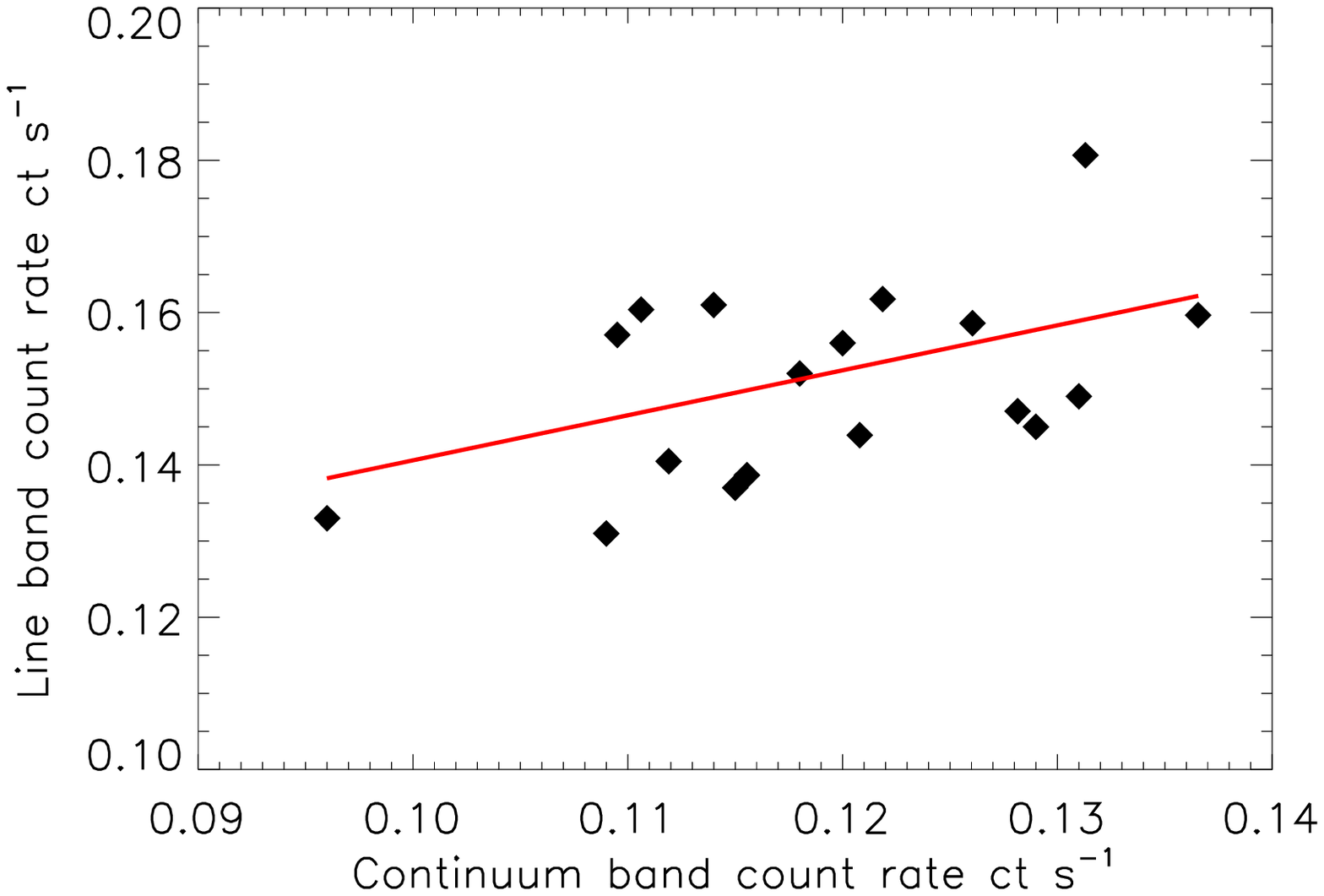}}
        \caption{Scatter plots for observations of the Groth-Westhall
          Strip. These observations were studied spectrally by
          \citet{kuntz2008} and cases $\{$13$\}$ and $\{$10$\}$ were
          found to have SWCX-enhancement, whereas $\{$64$\}$ was
          devoid of SWCX features.}
     \label{figgwsscat}
\end{figure*}

Several observations in our sample, used as control subjects, have
been previously studied by others and have been identified as having
experienced SWCX-enhancement. The \citet{snowden2004} observation of
the Hubble Deep Field North (case $\{7\}$, observation number
0111550401) produced a \redc\ value of 8.88 and is ranked highly
likely in Table \ref{tabresults}. \citet{kuntz2008} looked at several
observations of the Groth-Westfall Strip. These were (observation
numbers 0127921001(GWS1), 0127921101(GWS2) and 0127921201(GWS3)) and
the first two were reported to have enhanced OVII and OVIII emission
(correlated with enhanced solar wind flux as compared to a low solar
proton flux for GWS3). GWS1 (case $\{12\}$) and GWS2 (case $\{9\}$)
had \redc\ values of 4.7 and 5.67 whereas GWS3 gives a \redc\ value of
only 1.2 (case $\{63\}$, not included in Table \ref{tabresults}). GWS1
is judged to be very contaminated by residual soft protons, as is
mentioned in \citet{kuntz2008} and it showed considerable particle
contamination. The lightcurves for the line and continuum band along
with the ACE solar proton flux during the observation are plotted in
Figure \ref{figgwslcs}. Scatter plots for these observations are
plotted in Figure \ref{figgwsscat}. An observation of the Polaris
Flare region considered in the analysis of \citet{kuntz2008}, which
was reported to have a SWCX-enhancement, is ranked as case $\{31\}$
and therefore narrowly misses out on inclusion in Table
\ref{tabresults}. Other observations discussed by \citet{kuntz2008}
were ranked as unlikely to have experienced a SWCX-enhancement using
our criteria, which was in correlation with the diagnosis by
\citet{kuntz2008}, when no SWCX-enhancement was found. However, there
was one exception to this case where our indicators suggested no SWCX
effects had occurred whereas \cite{kuntz2008} states the contrary
(case $\{103$\}, observation number 0162160201). This is an example of
a false negative detection by our grading system whereby formally the
scatter of the line to continuum band count rates is insignificant and
therefore the observation is disregarded. Also, our criteria will not
identify observations that are affected by uniform SWCX throughout the
entirety of their exposure period, which seems to be the case in this
example. It should be noted in this particular case the number of
degrees of freedom was not particularly high and on inspection of the
lightcurves no obvious SWCX-enhanced period could be easily
identified.

\subsection{Cases of newly identified SWCX-enhancement}\label{seccasesall}
As shown in Table \ref{tabresults}, there is a large number of
observations that show high variability between the line and continuum
band and therefore produce a high \redc\ value. Index numbers in this
section refer to the index numbers of Table \ref{tabresults}. In this
section we briefly describe observations with currently unpublished
SWCX-enhancements, starting with those cases that have strong
indicators. The XMM-Newton lightcurves, orbital positions and spectra
for these cases are shown in Figure \ref{figresstrong}. We also
present several cases where SWCX-enhancements might have occurred and
which we classify as weak or dubious cases. The XMM-Newton orbital
positions, lightcurves and spectra for these cases are presented in
Figure \ref{figresweak}. Orbital positions are plotted in the
geocentric solar-ecliptic X-Y plane and the geocentric solar-ecliptic
X-Z plane. These plots are designed to aid the visualisation of the
line of sight of XMM-Newton during an observation; in particular we
wished to test whether the telescope was observing through the zone of
the brightest model geocoronal x-ray flux, as shown in
\citet{robertson2006}. The dark blue lines show the approximate
minimum and maximum positions, at the time of the observation, of the
magnetopause and bow shock respectively. At the subsolar point this is
calculated using \citet{khan1999} and the shape of these boundaries
have been approximated by a parabola. The path of XMM-Newton during
the period of the observation is shown in these plots and is coloured
black for the stages of the orbit which corresponds to the period of
expected SWCX-enhancement and red otherwise. The dotted lines on these
plots represent the line of sight of the telescope and the start and
end of the observation.

There are several cases which appear in Table \ref{tabresults}
  but that are not described below for a number of reasons.
\begin{enumerate}
\item The overall variation in both lightcurves was significantly
  large that a detection of variation in the line curve could not be
  made, resulting in a high value of \redc\ but a low value of
  \cratio\ (cases 8, 11, 15, 20, 23, 26, 27 and 29).
\item There was no time in the lightcurves when the line band tracked
  the continuum band meaning that \redc\ was still high, so if SWCX was
  present, it was spread throughout the observation (13, 16, 18, 19,
  21 and 22).
\item There was significant residual soft proton contamination after
  flare removal, meaning that it was statistically difficult to detect
  variation unique to the line band (24 and 30).
\end{enumerate}

\subsubsection{Example strong cases of SWCX-enhancement}\label{secswcxstrong}
\begin{itemize}
\item Case: $\{1\}$  0085150301 \\
  This observation shows a good correlation between the line band
  lightcurve and a peak in the solar proton flux. For the non-enhanced
  period, the line and continuum bands show the same count rate
  structure. This is the most extreme case of SWCX-enhancement in our
  sample and the richest in terms of line emission species
  observed. The OVIII clearly becomes visible during the enhancement
  period. In addition, many other emission lines are apparent. Below
  the oxygen lines, the carbon lines have appeared, but more
  strikingly clear are the NeIX and MgXI lines, at 0.91\,keV and
  1.34\,keV respectively. There are also emission lines seen between
  0.65\,keV and 1\,keV that could be attributed to OVIII or various
  species of iron, and an enhancement slightly above the silicon
  instrumental line at approximately 1.8\,keV. An increase in
  continuum level is seen during the SWCX period between 1100\,eV and
  1275\,eV. This observation shows considerable residual soft proton
  contamination, indicated by the high \fin\ value, which may explain
  the elevation of the continuum level. However, this does not explain
  the increase in ratio between a particular line and the continuum
  band, which we attribute to SWCX-enhancement. In Figure
  \ref{fig0085150301ds} we plot the resultant spectrum, using the non
  SWCX-enhanced period as the background spectrum. The background has
  been scaled to the SWCX-enhanced period between 2.1\,keV and
  2.5\,keV, to account for residual soft proton contamination
  throughout the observation. We fit the resulting spectrum with the
  lines detailed in Table \ref{tab0085150301lines}, giving the
  identification of the corresponding ion we consider responsible. It
  was necessary to add the instrumental fluorescent lines Al-K$_{\alpha}$
  and Si-K$_{\alpha}$ to the fit. This background component is time
  variable and therefore is not completely accounted for by the
  scaling between the non-enhanced and SWCX-enhanced period, hence
  this residual component is still present in the spectrum.  We also
  consider the possibility that this enhancement is linked to the CME
  event of the 19th October 2001 \citep{wang2005}. The delay between
  the occurrence of this event at the solar corona and its arrival near
  Earth would be approximately three days. This CME was registered by
  ACE and therefore passed in the relatively near vicinity of the
  Earth. Such an event could explain the large enhancements and
  richness of the spectrum, seen for this example of SWCX-enhancement.

\begin{table}
  \caption{SWCX line emission seen in the spectrum of case \{1\}.}
\begin{tabular}{|c|c|}
\hline
Ion    & Line energy (keV)   \\ 
\hline
CVI    & 0.37  \\
CVI    & 0.46  \\
OVII   & 0.56  \\
OVIII  & 0.65   \\
OVIII/FeXVII? & 0.81   \\
NeIX   & 0.91   \\
NeX/FeXVII?  & 1.03   \\
NeIX   & 1.15  \\
FeXVII & 1.20 \\
MgXI   & 1.34  \\
MgXII/SiXIII?& 1.85  \\
SXIV ?& 2.000 \\
\hline
\end{tabular}
\label{tab0085150301lines}
\end{table}

\begin{figure}[h]
      \includegraphics[width=.33\textwidth, angle=270]{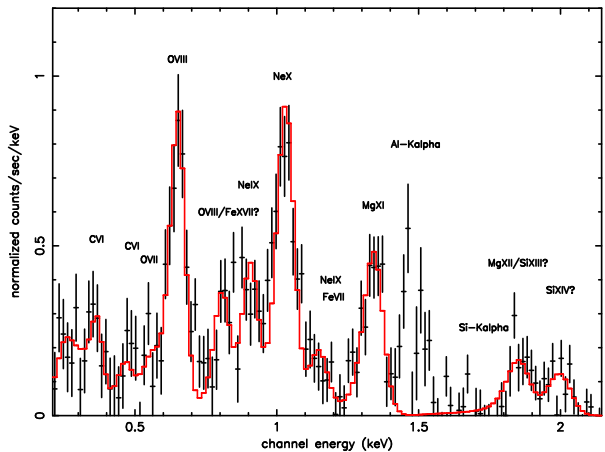}
      \caption{The resultant spectrum for case \{1\}. The model to the
        SWCX lines folded through the instrument response is plotted
        in red. The excess seen around 1.49\,keV and 1.74\,keV results
        from particle-induced instrumental background.}
      \label{fig0085150301ds}
\end{figure}

\item Case: $\{3\}$ 0149630301 \\
  This observation towards the Large Magellanic Cloud shows a line
  band over continuum band enhancement in the latter part of the
  observation. ACE also shows a solar proton flare that rises in the
  second half of the observation. XMM-Newton is travelling in the nose
  of the magnetosheath. The enhancement was expected in the latter
  part of the observation, approximately two hours after the start of
  the observation. The \crh\ shows a good Gaussian fit and the \fin\
  ratio is low. The background time series file however, shows a large
  flare at the end of the observation that adds some doubt to our
  diagnosis. The \cratio, at 9.47, is high. The spectrum created for
  the supposed enhanced period is interesting as it shows low energy
  enhancement below 0.7\,keV along with an apparent line at around
  1.35\,keV that could be attributed to MgXI.
\item Case: $\{4\}$  0305920601 \\
  This observation was highly variable in the oxygen band, producing a
  very high \redc\ value of approximately 15. It had a low value for
  the \fin\ ratio and a good fit to the count rate histogram from the
  ESAS software. The \cratio\ was the highest in the sample. This
  observation was conducted in the summer months, so XMM-Newton was
  found in the nose of the magnetosheath. The solar wind flux was
  relatively high throughout the observation, with a peak greater than
  1$\times$10$^{9}$\,cm$^{-2}$\,s$^{-1}$. It can be observed from the
  lightcurve that there is a step at approximately two hours, before
  which the line band shows much more variability than the continuum
  band. The enhancement period shows increased flux from the OVIII
  line and below and possibly some increase around 0.9\,keV which
  could be attributed to NeIX.
\item Case:  $\{5\}$  0150680101 \\
  This observation was conducted in July so that XMM-Newton was
  positioned in the nose of the magnetosheath. The ACE data are a
  little sparse, but when available the solar proton flux was high,
  peaking above 1.5$\times$10$^{9}$\,cm$^{-2}$\,s$^{-1}$. A time cut
  was taken at approximately nine hours as a SWCX-enhanced period was
  expected in the latter half of the observation. The spectrum shows
  OVII and OVIII enhancements and there is also a strong enhancement
  seen below 0.5\,keV where lines from carbon are expected.
\item Case: $\{14\}$  0136000101 \\
  The lightcurves for this observation showed a sharp peak in the line
  band during the last bin. This observation was taken in April when
  XMM-Newton is in the flanks of the magnetosheath at approximately 90
  degrees to the Earth-Sun line. The solar proton flux shows a sudden
  step to a particularly high level nearing
  2$\times$10$^{9}$\,cm$^{-2}$\,s$^{-1}$. Depending on the delay taken
  between ACE and the magnetosheath, which will be approximately one
  hour, this step could correspond to the peak seen in the
  lightcurves. The \fin\ ratio shows this observation to be very
  contaminated by residual soft protons. This case also shows an
  increase in flux from the OVIII line and below, although there is no
  evidence and very poor statistics for SWCX lines with greater
  energies, such as NeIX and MgXI.
\end{itemize}

\subsubsection{Example weak cases of SWCX-enhancement}\label{secswcxweak}

\begin{itemize}
\item Case: $\{2\}$ 0093552701 \\
  This observation was very contaminated by soft protons yet shows a
  high \cratio\ of 6.09. The combined spectrum for the enhanced period showed an
  increase in flux for both the OVII and OVIII lines over the
  non-enhanced spectrum, although it exhibits less lower energy flux
  around the carbon lines in comparison to the strong cases of
  SWCX-enhancement. There was little solar proton flux data available
  during this observation.
\item Case: $\{6\}$ 0101040301 \\
  This observation shows a very high \cratio\ of 4.06 and is very
  contaminated by residual soft protons. The slow rise in the solar
  proton flux could be correlated with the rise in the line band
  during the latter part of the observation. XMM-Newton was observing
  through the flanks of the magnetosheath throughout. The spectral
  case for SWCX-enhancement however is not as strong as for other
  cases.
\item Case: $\{10\}$ 0070340501 \\
  This observation is quite possibly affected by SWCX but has very few
  bins left after GTI-filtering and so we are reluctant to make any
  strong conclusions regarding the diagnosis. Emission lines of OVII,
  OVIII and some carbon are seen for the suspected enhanced
  period. The solar proton flux for this observation showed a peak
  that could correspond to the peak seen in the line band counts,
  although the increase in counts is seen directly after a filtered
  period of the lightcurve.
\item Case: $\{17\}$ 0101440101 \\
  This observation has very high levels of residual soft proton
  contamination and has been heavily GTI-filtered, but the spectra for
  the SWCX period show an enhancement below 0.7\,keV suggestive of
  SWCX-lines without any increase in flux for either the enhanced and
  non-enhanced periods above this energy. The enhanced period was
  taken after 12 hours. The \cratio of 1.56 is quite high. No
  correlation between the line band lightcurve and solar proton flux
  is seen, although XMM-Newton is positioned on the sub-solar side of
  the magnetosheath throughout the observation. For the enhancement
  period both the OVII and OVIII become apparent and there is possibly
  some increase in flux around the carbon lines below 0.5\,keV.
\item Case: $\{25\}$ 0164560701 \\
  This observation is contaminated by residual soft protons. It does
  not have a perfect fit to the count rate histogram from the ESAS
  software and not a particularly high \cratio\ of 1.53. The spectrum
  for the enhanced period shows very low level statistics although
  there is evidence of enhancement around the oxygen and carbon
  lines. The spectral split was taken at a time of seven hours. The
  observation was conducted in July when XMM-Newton was in the nose of
  the magnetosheath.
\item Case: $\{28\}$ 0106460101 \\
  This observation was taken in November when XMM-Newton was looking
  though the flanks of the magnetosheath. There is possibly some
  correlation between the line band lightcurve and the solar proton
  flux. The period for possible SWCX-enhancement was chosen to be
  between approximately two and seven hours. OVII, OVIII and some
  carbon enhancement is seen.
\end{itemize}


\begin{figure*}[]
     \subfigure[Case: $\{$1$\}$, obsn. 0085150301]{\label{fig0085150301}\includegraphics[width=\textwidth, bb=0 640 560 730, clip=]{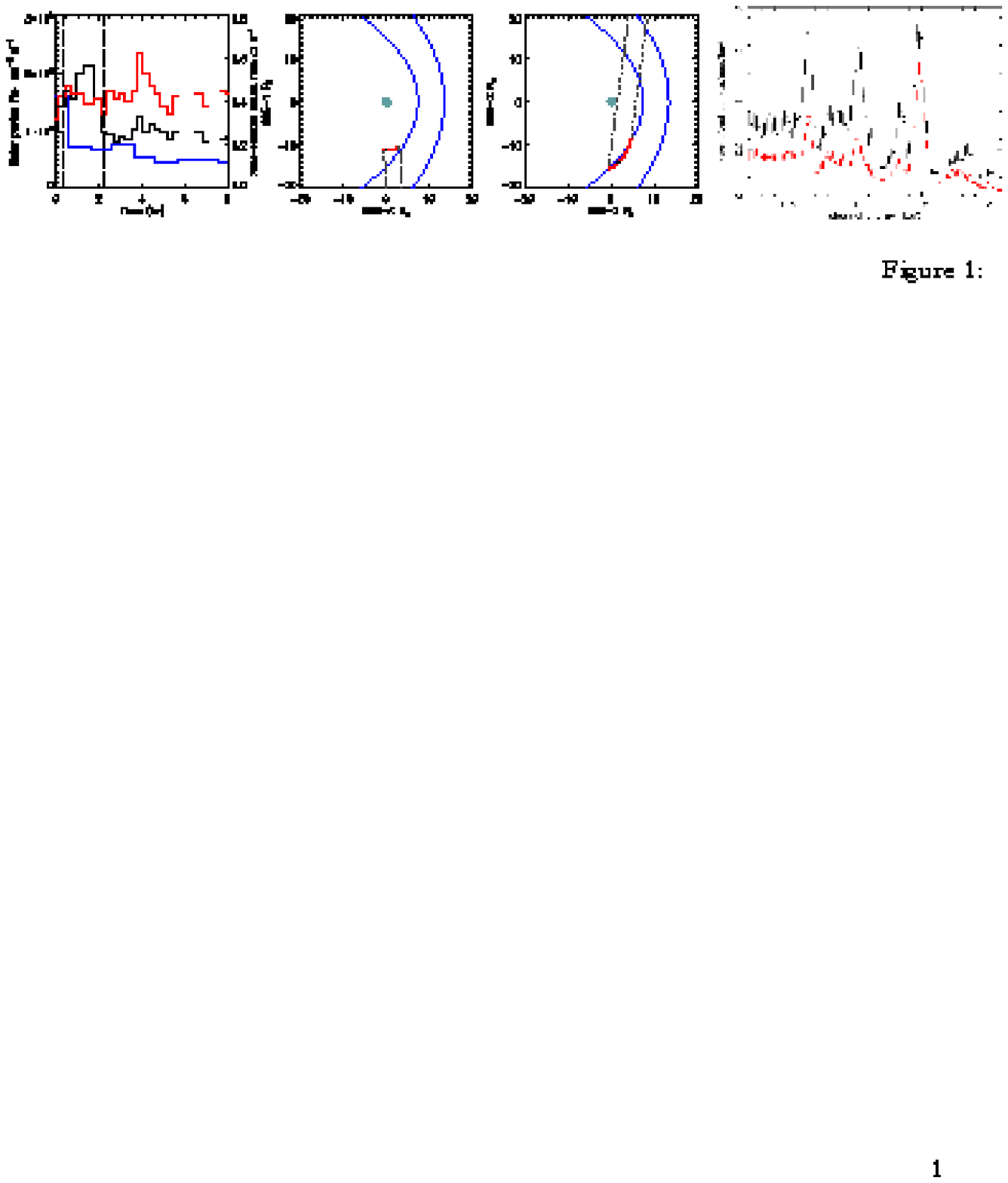}}
     \subfigure[Case: $\{$3$\}$, obsn. 0149630301]{\label{fig0149630301}\includegraphics[width=\textwidth, bb=0 640 560 730, clip=]{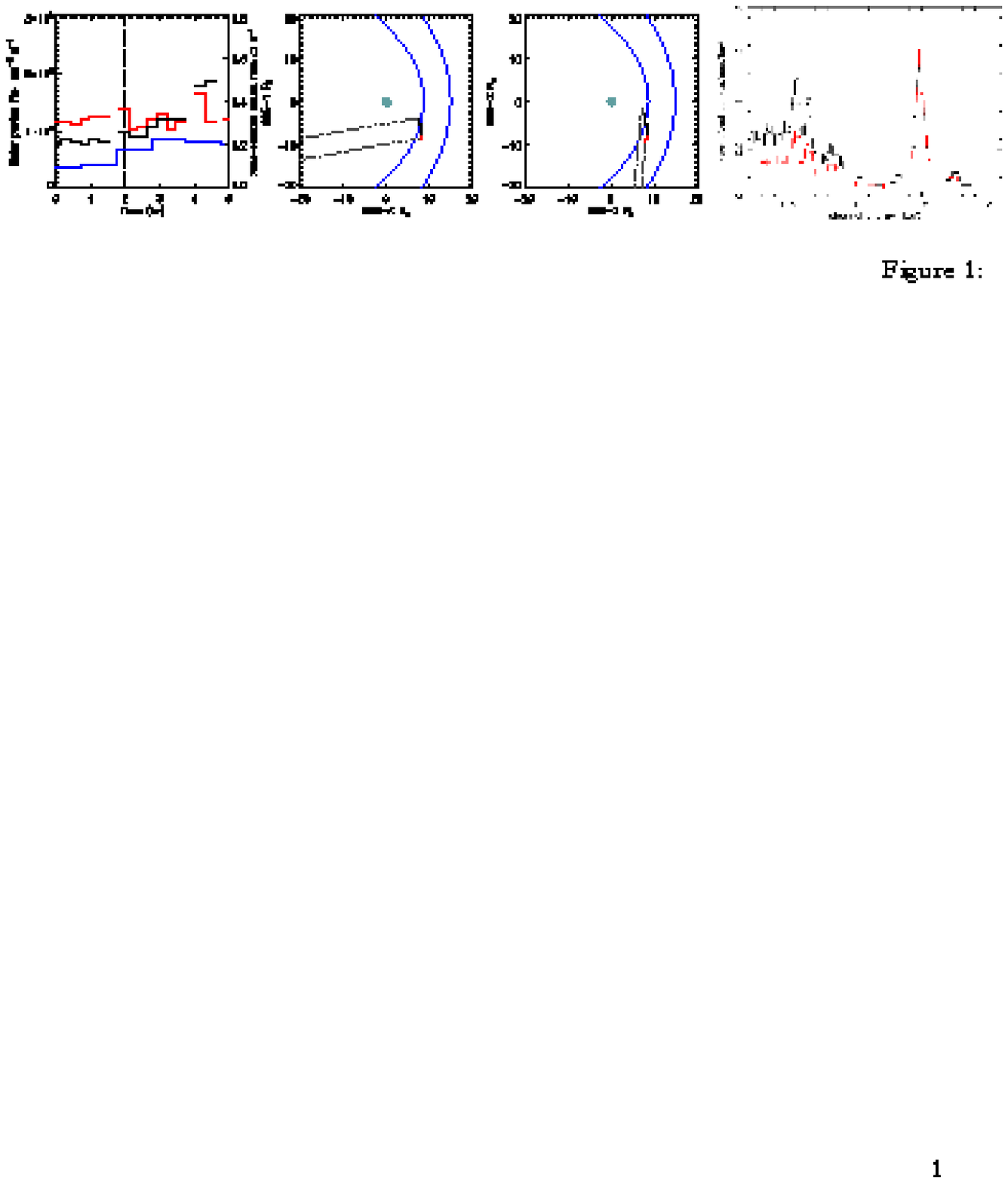}}
     \subfigure[Case: $\{$4$\}$, obsn. 0305920601]{\label{fig0305920601}\includegraphics[width=\textwidth, bb=0 640 560 730, clip=]{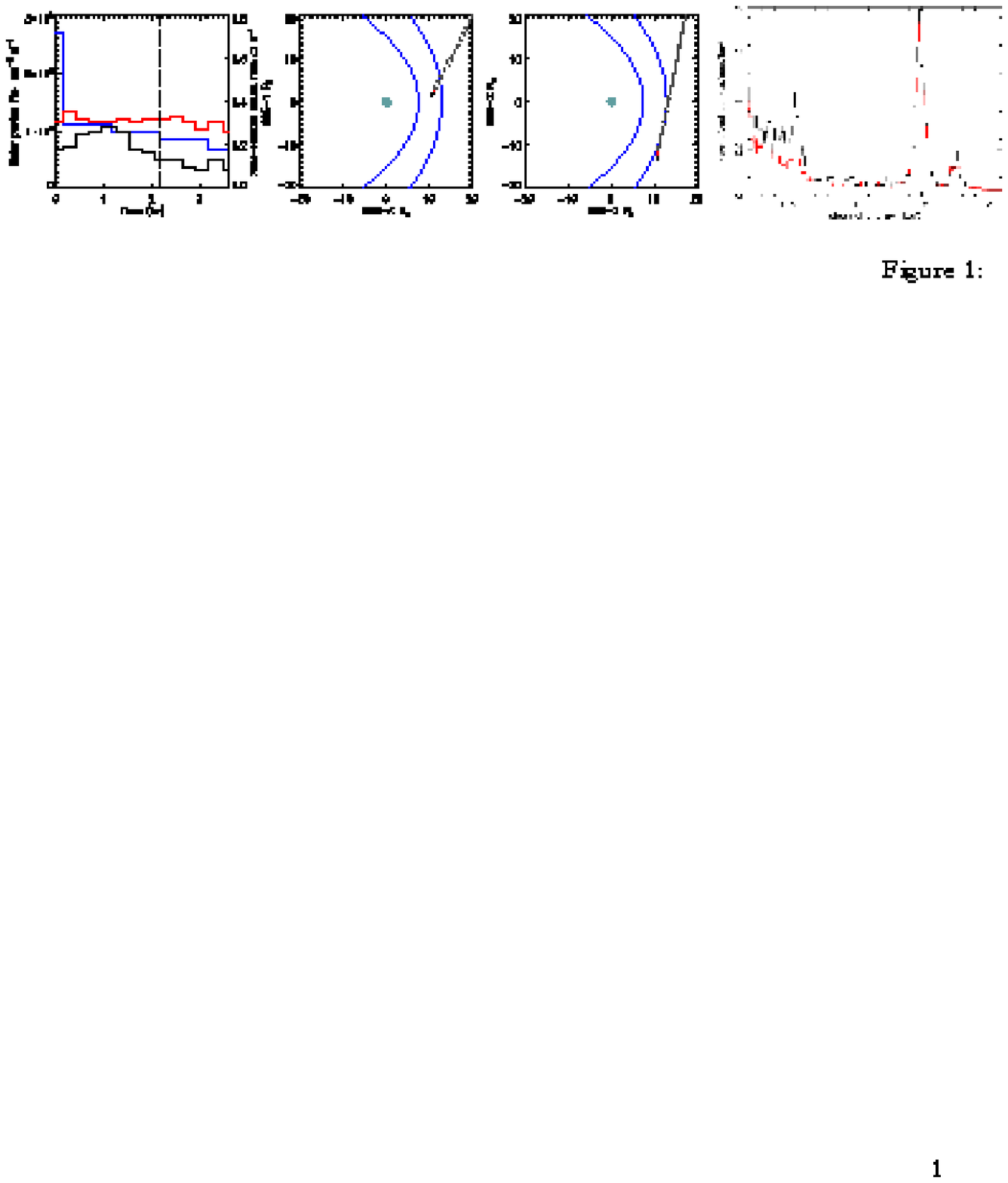}}
     \subfigure[Case: $\{$5$\}$, obsn. 0150680101]{\label{fig0150680101}\includegraphics[width=\textwidth, bb=0 640 560 730, clip=]{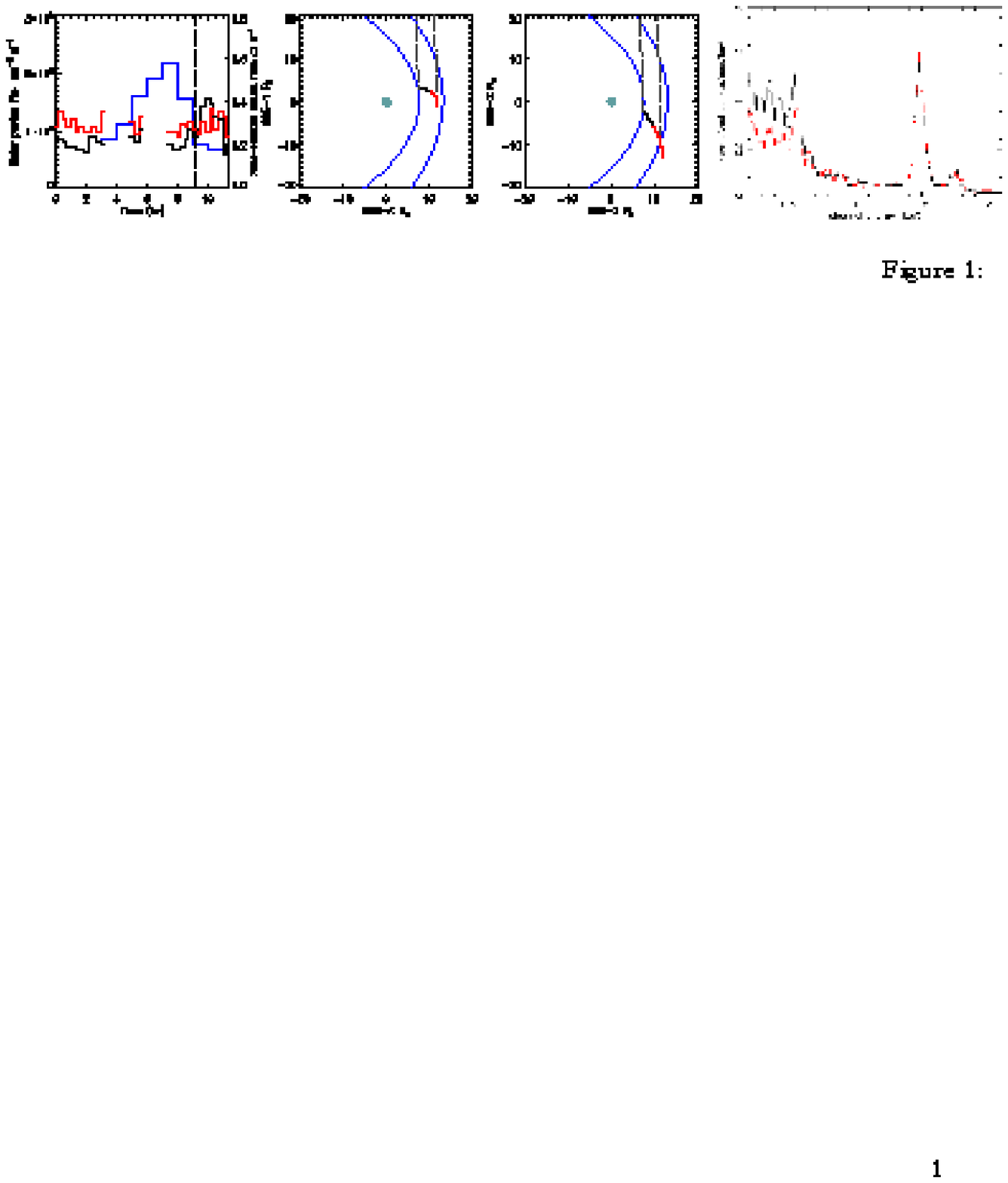}}
     \subfigure[Case: $\{$14$\}$, obsn. 0136000101]{\label{fig0136000101}\includegraphics[width=\textwidth, bb=0 640 560 730, clip=]{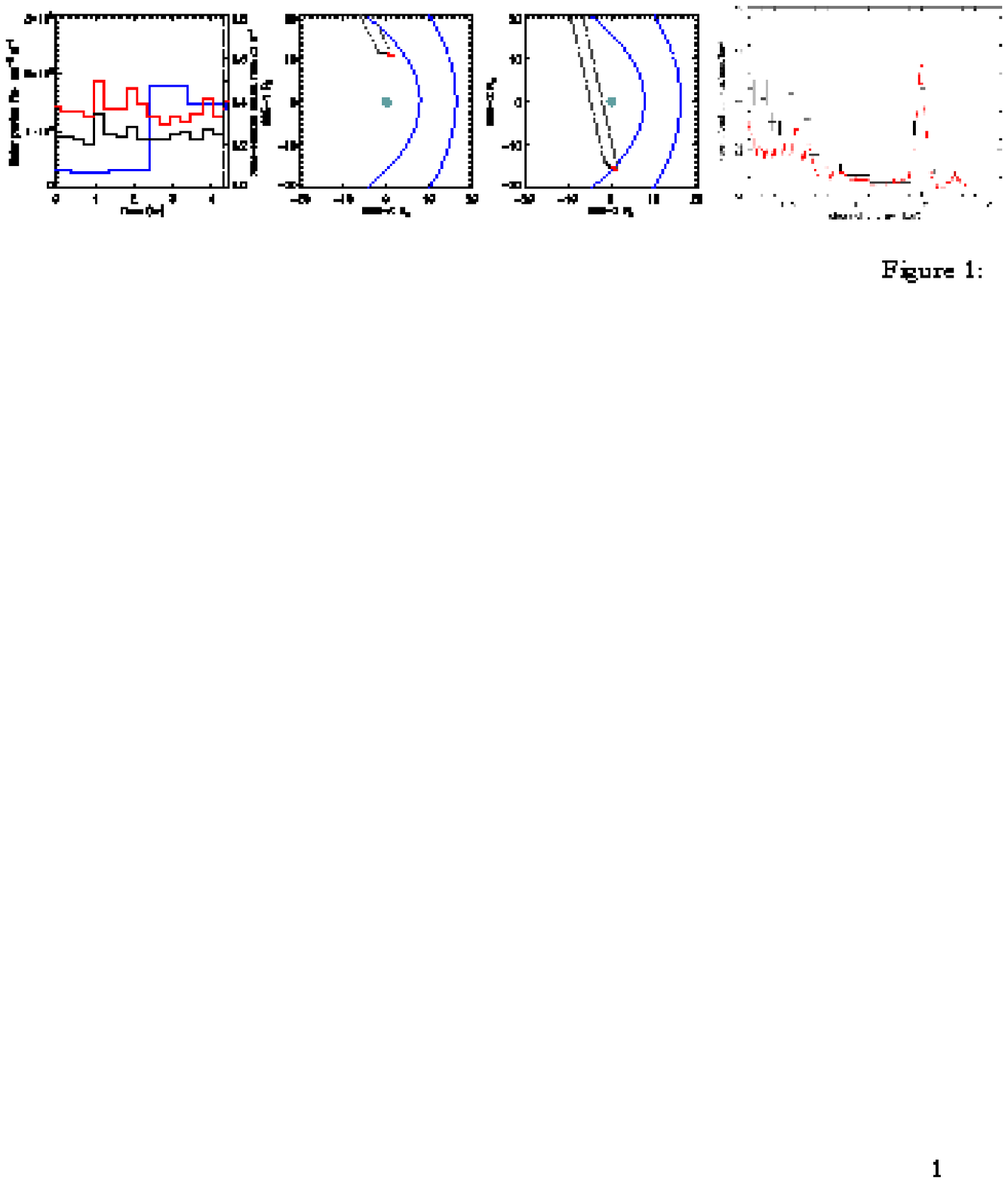}}
     \caption{XMM-Newton lightcurves, orbital positions and spectra
       for example cases of SWCX-enhancement. In the first panel for
       each case, the blue line represents the solar proton flux
       during each observation, plotted including the expected one
       hour delay and the vertical lines indicate the start or stop
       times for the division of the events into SWCX-enhanced and non
       SWCX-enhanced periods. Black and red represent the line and
       continuum band lightcurves respectively, similarly for the
       spectra shown in the last panel. The second and third panels
       show XMM-Newton in orbit around the Earth in the GSE X-Y and
       X-Z planes. The positive x-axis is directed towards the Sun and
       the positive y-axis is directed in the direction of the orbit
       of XMM-Newton. Full resolution
       images can be found at www.star.le.ac.uk/$\sim$jac48/publications/var\_swcx\_aa.pdf}
     \label{figresstrong}
\end{figure*}

\begin{figure*}[]
     \centering
     \subfigure[Case: $\{$2$\}$, obsn. 0093552701]{\label{fig0093552701}\includegraphics[width=\textwidth, bb=0 640 560 730, clip=]{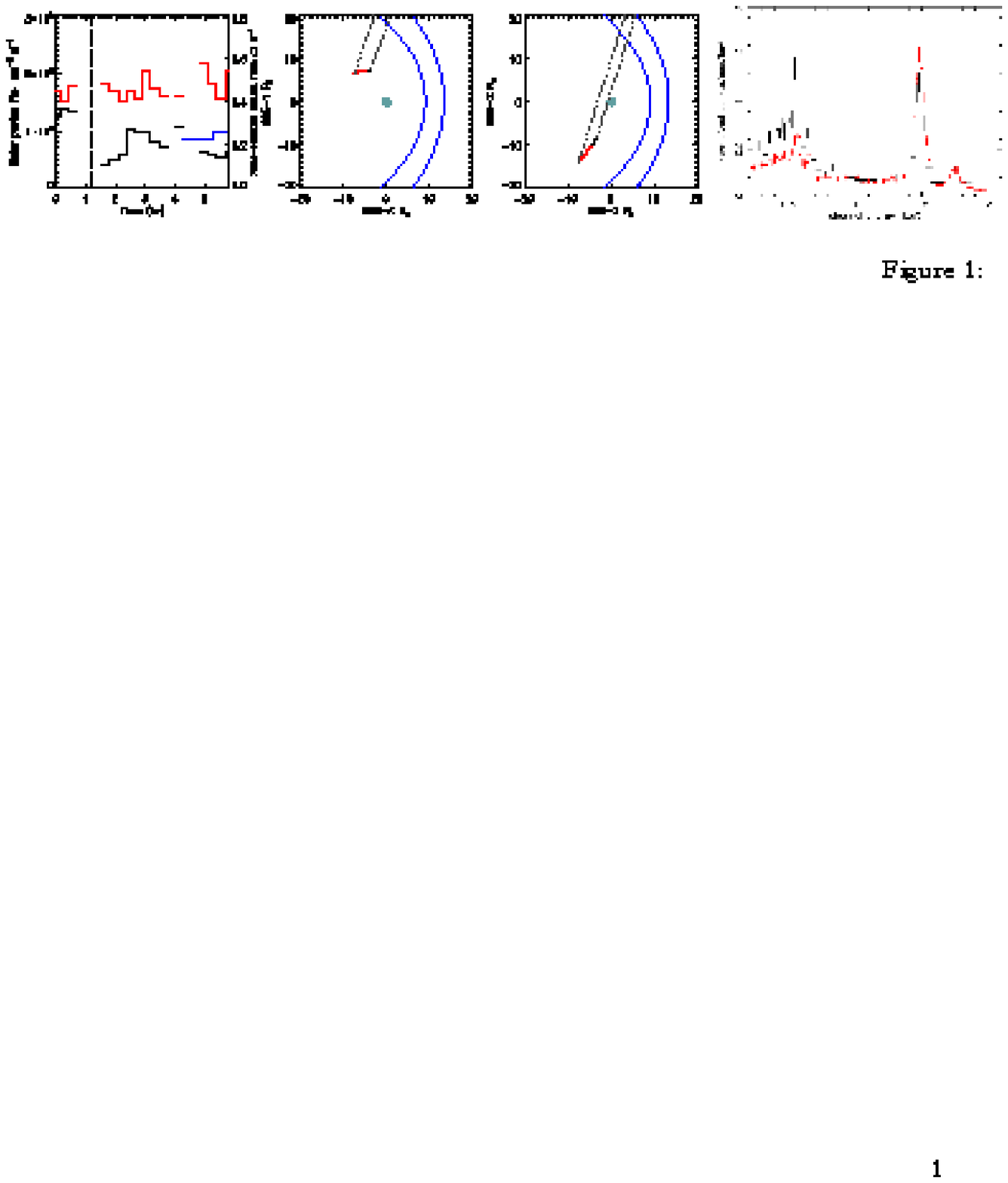}} 
     \subfigure[Case: $\{$6$\}$, obsn. 0101040301]{\label{fig0101040301}\includegraphics[width=\textwidth, bb=0 640 560 730, clip=]{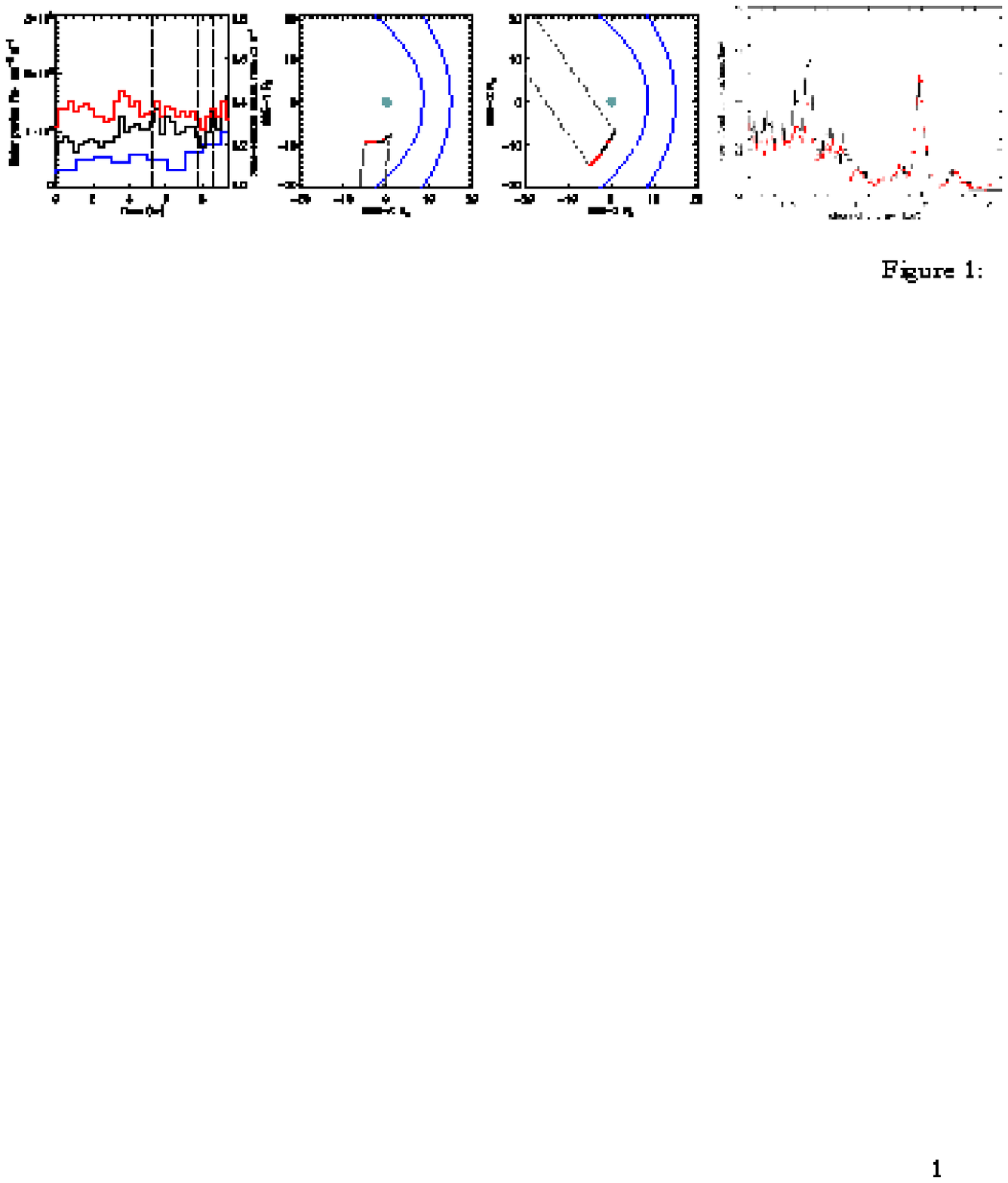}}
     \subfigure[Case: $\{$10$\}$, obsn. 0070340501]{\label{fig0070340501}\includegraphics[width=\textwidth, bb=0 640 560 730, clip=]{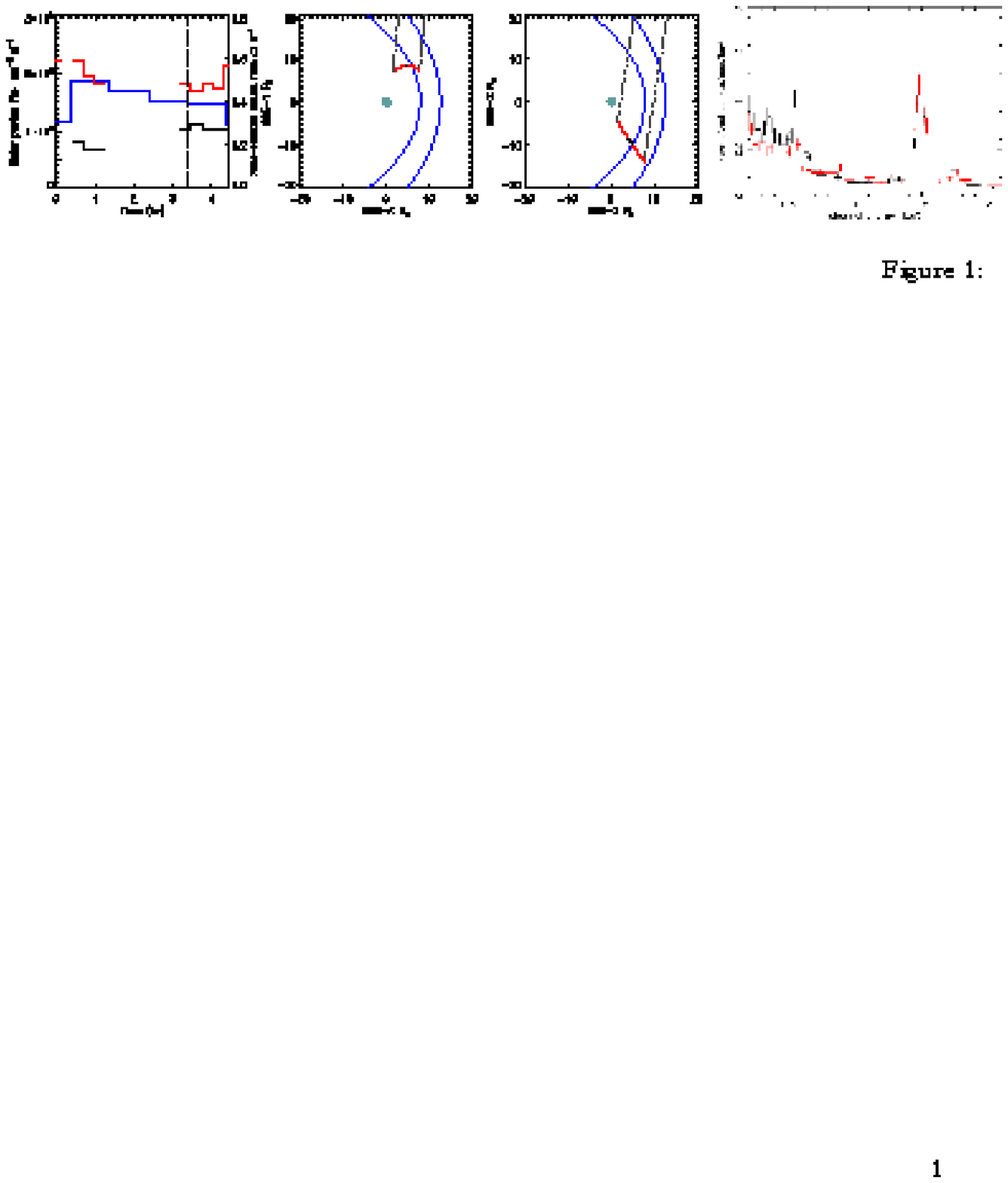}}
     \subfigure[Case: $\{$17$\}$, obsn. 0101440101]{\label{fig0101440101}\includegraphics[width=\textwidth, bb=0 640 560 730, clip=]{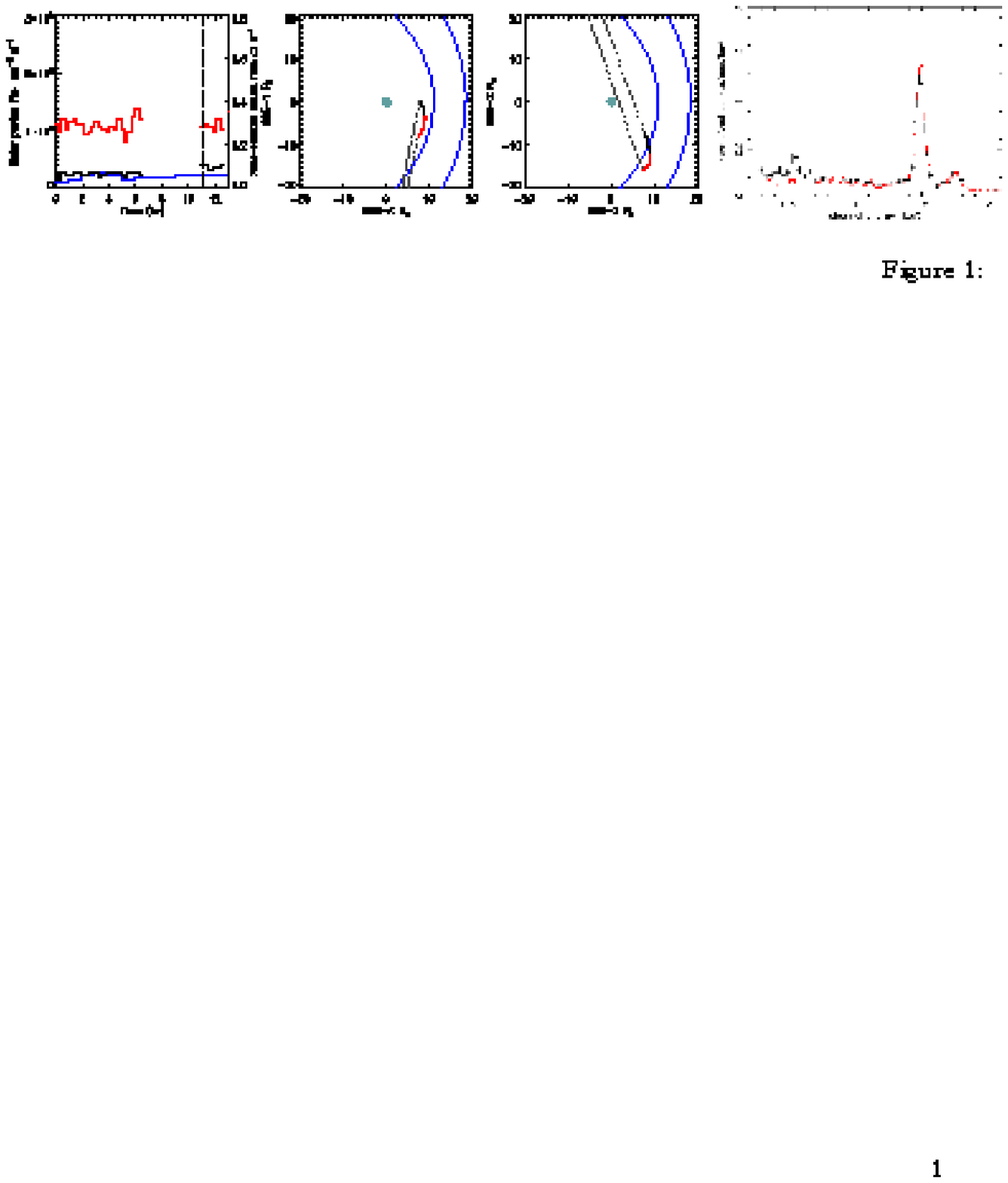}}
     \subfigure[Case: $\{$25$\}$, obsn. 0164560701]{\label{fig0164560701}\includegraphics[width=\textwidth, bb=0 640 560 730, clip=]{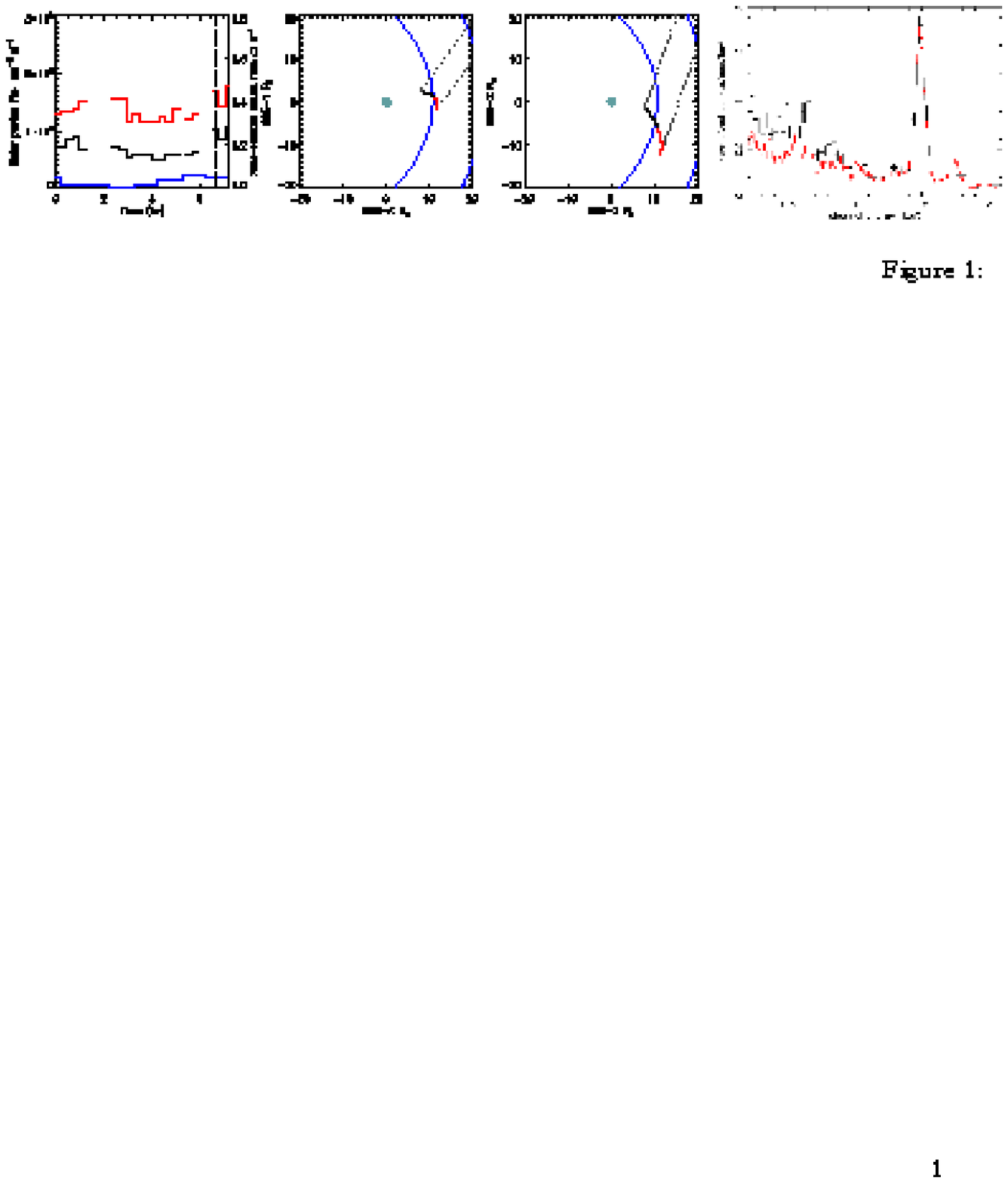}} 
     \subfigure[Case: $\{$28$\}$, obsn. 0106460101]{\label{fig0106460101}\includegraphics[width=\textwidth, bb=0 640 560 730, clip=]{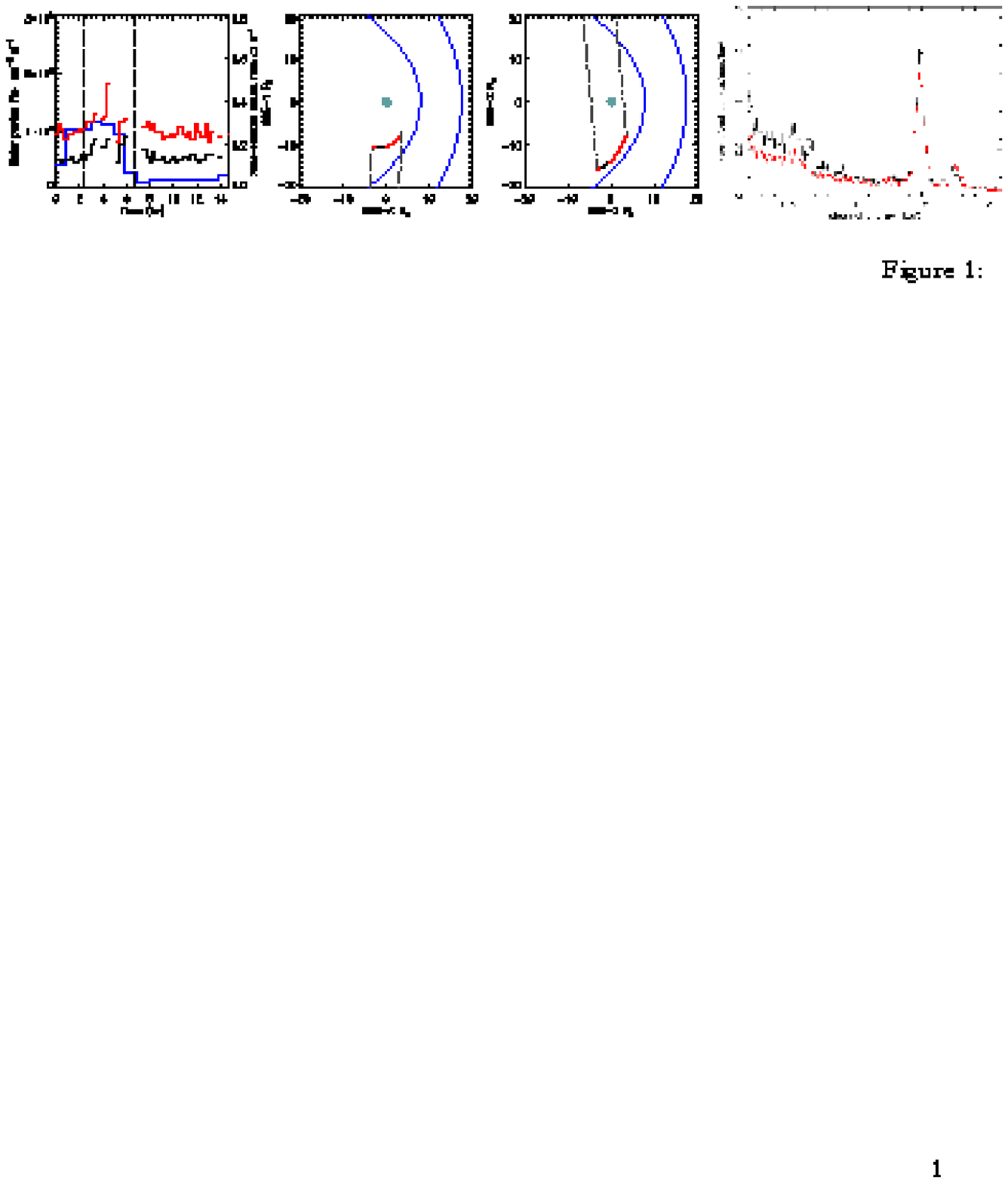}} 
     \caption{Lightcurves, orbital positions and spectra for suspected
       cases of weak or dubious SWCX-enhancement, arranged in panels
       as for the strong cases of SWCX-enhancement. Full resolution
       images can be found at www.star.le.ac.uk/$\sim$jac48/publications/var\_swcx\_aa.pdf}
     \label{figresweak}
\end{figure*}

\section{Conclusions and further work}\label{secconcl}
We have presented a selection of results from a preliminary sample of
approximately 170 XMM-Newton observations that have been tested for
oxygen line variability as compared to a continuum band that is not
expected to vary. Our primary measure of the likelihood of
SWCX-enhancement is found through the lack of correlation between a
count rate for the prime SWCX indicators, namely the band representing
the OVII and OVIII major emission lines, and the count rate for a band
that is SWCX-enhanced emission free. SWCX-enhanced observations showed
both a high \redc\ and \cratio\ value. In the event of soft proton
contamination, correlation between the count rates from both bands is
still expected. The level of anti-correlation is assessed using the
metric of the \redc\ from the relationship between a band representing
the SWCX indicators and a band representing the emission line free
continuum. We have demonstrated that the method has been successful in
identifying cases of SWCX from a medium sized sample, providing
careful consideration of various other diagnostic tests are taken into
account. In addition, our method was also able to identify all but one
of the control observations with previously identified SWCX that had
been placed in the sample, within the top 35 \redc-ranked results.

Our strongest case of SWCX-enhancement, with the richest spectrum
showing many emission lines, gave the highest level of the \redc\
value and presented a high \cratio. This observation was contaminated
by residual soft protons, although the resultant spectrum, between the
non and SWCX-enhanced periods, was able to show many emission lines
corresponding to highly charged ion species.

Cases which we believe present the best evidence of SWCX-enhancement
exhibit enhanced spectral features such as emission lines of OVII and
OVIII, along with lower energy lines probably associated with
carbon. Two of the strongest cases offer emission lines that may be
attributed to NeIX and MgXI. The pattern of observed emission lines
depends on the exact composition of the solar wind at 1AU at the time
of the observation.

We have shown that several of the cases where SWCX-enhancements are
suspected occurred during periods when XMM-Newton was found to be on
the sub-solar side of the magnetosheath. The line of sight of the
telescope in these cases intersects the area of the magnetosheath of
brightest expected solar wind-induced geocoronal X-ray flux. The
SWCX-enhancement seen for other cases, when XMM-Newton was found to
have a line of sight which intersects the magnetosheath through the
weaker areas of X-ray flux, is probably due to a non-geocoronal
origin, such as a CME density enhancement passing through the
heliosphere.

In Paper II we will present results from the processing of a larger
sample of observations to identify additional observations with
SWCX-enhancement, extending our analysis to include data from the PN
camera. We expect to find approximately 2000 observations of
XMM-Newton affected by temporally variable SWCX, based on frequency of
detections from the sample described in this paper. However, the
number of detections is dependent on solar activity and therefore may
vary considerably from the number stated. Using a larger sample, we
aim to refine the identification of SWCX-enhancement, using the
relationship between \redc\ and \cratio.

We will include an improved algorithm to remove sources in the field,
reducing the danger of residual source contamination. We will
investigate for any trends within the results that are related to the
solar cycle and the position angle and orientation of XMM-Newton
during each observation. This will help us to constrain the conditions
when XMM-Newton is most likely to experience SWCX, whether that be due
to the satellite's position or viewing geometry. We will also explore
multiple pointings of the same target to consider variations between
observations. We will investigate the observed SWCX line ratios and
what this implies for the composition of the solar wind.

\section{Acknowledgements}\label{secackno}
We would like to thank Simon Vaughan for helpful discussions regarding
timing analysis and statistics and Kip Kuntz for similarly helpful
discussions and advice. We would also like to thank the referee for
his insightful comments and suggestions. This work has been funded by
the Science and Technology Facilities Council, U.K.

\bibliographystyle{aa} 
\bibliography{phd_gen_aa} 

\end{document}